\documentclass[a4paper]{article}

\usepackage{fullpage}
\usepackage[english]{babel}
\usepackage{setspace}
\usepackage[square]{natbib}
\usepackage{endfloat}
\usepackage{geometry}
\usepackage{multirow}
\usepackage{amsmath}
\usepackage{amssymb}
\usepackage{graphicx}
\usepackage{lineno}
\usepackage{rotating}
\usepackage{hyperref}

\linenumbers*[1]

\hypersetup{breaklinks=true,
  pdftitle={Modeling All Exceedances Above a Threshold Using an
    Extremal Dependence Structure: Inferences on Several Flood
    Characteristics},
  pdfkeywords={Extreme Value Theory, Pickands' Dependence Function,
    Markov Chains, Extreme Value Dependence Structure},
  pdfauthor={Mathieu Ribatet}
}
\title{Modeling All Exceedances Above a Threshold Using an Extremal
  Dependence Structure:\\ Inferences on Several Flood
  Characteristics}
\date{}

\author{Mathieu Ribatet$^{\ast,\dag}$ \and Taha B.M.J. Ouarda$^\dag$
  \and Eric Sauquet$^\ast$ \and Jean-Michel Gr\'esillon$^\ast$}

\begin{document}
\maketitle

\begin{center}
  Submitted to: \textit{Water Resources Research}

  \vspace{0.2cm}
  $^\ast$ Cemagref Lyon, Unit\'e de Recherche Hydrologie-Hydraulique, 3
  bis quai Chauveau, CP220, 69336 Lyon cedex 09, France 

  $^\dag$ INRS-ETE, University of Qu\'ebec, 490, de la Couronne
  Qu\'ebec, Qc, G1K 9A9, CANADA. 

  \vspace{0.2cm}
  Corresponding author: M. Ribatet; Email:  
  \href{mailto:ribatet@lyon.cemagref.fr}{ribatet@lyon.cemagref.fr}\\
  Phone: +33 4 72 20 87 64; Fax: +33 4 78 47 78 75
\end{center}
\begin{doublespace}
  \begin{abstract}
    \noindent
    Flood quantile estimation is of great importance for many engineering studies and policy decisions. However, practitioners must often deal with small data available. Thus, the information must be used optimally. In the last decades, to reduce the waste of data, inferential methodology has evolved from annual maxima modeling to peaks over a threshold one. To mitigate the lack of data, peaks over a threshold are sometimes combined with additional information - mostly regional and historical information. However, whatever the extra information is, the most precious information for the practitioner is found at the target site. In this study, a model that allows inferences on the whole time series is introduced. In particular, the proposed model takes into account the dependence between successive extreme observations using an appropriate extremal dependence structure. Results show that this model leads to more accurate flood peak quantile estimates than conventional estimators. In addition, as the time dependence is taken into account, inferences on other flood characteristics can be performed. An illustration is given on flood duration. Our analysis shows that the accuracy of the proposed models to estimate the flood duration is related to specific catchment characteristics.  Some suggestions to increase the flood duration predictions are introduced.
  \end{abstract}

  \section{Introduction}
  \label{sec:intro}

  Estimation of extreme flood events is an important stage for many
  engineering designs and risk management. This is a considerable task
  as the amount of data available is often small. Thus, to increase
  the precision and the quality of the estimates, several authors use
  extra information in addition to the target site one. For example,
  \citet{Ribatet2007}, \citet{Kjeldsen2006,Kjeldsen2007} and
  \citet{Cunderlik2006b} add information from other homogeneous gaging
  stations. \citet{Werritty2006} and \citet{Reis2005} use historical
  information to improve inferences.  Incorporation of extra
  information in the estimation procedure is attractive but it should
  not be more prominent than the original data \citep{Ribatet2007c}.
  Before looking at other kinds of information, it seems reasonable to
  use efficiently the one available at the target site. Most often,
  practitioners have initially the whole time series, not only the
  extreme observations.  In particular, it is a considerable waste of
  information to reduce a time series to a sample of Annual Maxima
  (\textbf{AM}).

  In this perspective, the Peaks Over Threshold (\textbf{POT})
  approach is less wasteful as more than one event per year could be
  inferred. However, the declustering method used to identify
  independent events is quite subjective. Furthermore, even though a
  ``quasi automatic'' procedure was recently introduced by
  \citet{Ferro2003}, there is still a waste of information as only
  cluster maxima are used.

  \citet{Coles1994a} and \citet{Smith1997} propose an approach using
  Markov chain models that uses all exceedances and accounts for
  temporal dependence between successive observations. Finally, the
  entire information available within the time series is taken into
  account. More recently, \citet{Fawcett2006} give an illustrative
  application of the Markov chain model to extreme wind speed
  modeling.

  In this study, extreme flood events are of interest. The performance
  of the Markov chain model is compared to the conventional POT
  approach. The data analyzed consist of a collection of 50 French
  gaging stations.  The area under study ranges from 2$^\circ$W to
  7$^\circ$E and from 45$^\circ$N to 51$^\circ$N. The drainage areas
  vary from 72 to 38300 km$^2$ with a median value of 792 km$^2$.
  Daily observations were recorded from 39 to 105 years, with a mean
  value of 60 years. For the remainder of this article, the quantile
  benchmark values are derived from the maximum likelihood estimates
  on the whole times series using a conventional POT analysis.

  The paper is organized as follows.
  Section~\ref{sec:markov-chain-model} introduces the theoretical
  aspects for the Markov chain model, while
  Section~\ref{sec:ExtValDepTest} checks the relevance of the
  Markovian model hypothesis. Section~\ref{sec:perfMark}
  and~\ref{sec:infFlood} analyze the performance of the Markovian
  model to estimate the flood peaks and durations
  respectively. Finally, some conclusions and perspectives are drawn
  in Section~\ref{sec:conclusion}.

  \section{A Markov Chain Model for Cluster Exceedances}
  \label{sec:markov-chain-model}

  In this section, the extremal Markov chain model is presented. In the
  remainder of this article, it is assumed that the flow $Y_t$ at time
  $t$ depends on the value $Y_{t-1}$ at time $t-1$. The dependence
  between two consecutive observations is modeled by a first order
  Markov chain. Before introducing the theoretical aspects of the model,
  it is worth justifying and describing the main advantages of the
  proposed approach.

  It is now well-known that the univariate Extreme Value Theory
  (\textbf{EVT}) is relevant when modeling either AM or POT\@.
  Nevertheless, its extension to the multivariate case is surprisingly
  rarely applied in practice. This work aims to motivate the use of
  the Multivariate EVT (\textbf{MEVT}). In our application, the
  multivariate results are used to model the dependence between a set
  of lagged values in a times series. Consequently, compared to the AM
  or the POT approaches, the amount of observations used in the
  inference procedure is clearly larger. For instance, while only
  cluster maxima are used in a POT analysis, all exceedances are
  inferred using a Markovian model.
  
  \subsection{Likelihood function}
  \label{subsec:likfun}

  Let $Y_1,\ldots,Y_n$ be a stationary first-order Markov chain with a
  joint distribution function of two consecutive observations $F(y_1,
  y_2)$, and $F(y)$ its marginal distribution. Thus, the likelihood
  function $L$ evaluated at points $(y_1, \ldots, y_n)$ is:

  \begin{equation}
    \label{eq:likMC}
    L(y_1, \ldots, y_n) = f(y_1) \prod_{i=2}^n f(y_i | y_{i-1}) =
    \frac{\prod_{i=2}^n f(y_i, y_{i-1})}{\prod_{i=2}^{n-1} f(y_i)}
  \end{equation}
  where $f(y_i)$ is the marginal density, $f(y_i | y_{i-1})$ is the
  conditional density, and $f(y_i, y_{i-1})$ is the joint density of
  two consecutive observations.

  To model all exceedances above a sufficiently large threshold $u$,
  the joint and marginal densities must be known. Standard univariate
  EVT arguments \citep{Coles2001} justify the use of a Generalized
  Pareto Distribution (\textbf{GPD}) for $f(y_i)$ - e.g. a term of the
  denominator in equation~\eqref{eq:likMC}. As a consequence, the
  marginal distribution is defined by:

  \begin{equation}
    \label{eq:margDist}
    F(y) = 1 - \lambda \left(1 + \xi \frac{y - u}{\sigma}
    \right)_+^{-1/\xi}, \qquad y \geq u\\
  \end{equation}
  where $x_+ = \max(0, x)$, $\lambda = \Pr[Y \geq u]$, $\sigma$ and
  $\xi$ are the scale and shape parameters respectively.  Similarly,
  MEVT arguments \citep{Resnick1987} argue for a bivariate extreme
  value distribution for $f(y_i, y_{i-1})$ - e.g. a term of the
  numerator in equation~\eqref{eq:likMC}. Thus, the joint distribution
  is defined by:

  \begin{equation}
    \label{eq:jointDist}
    F(y_1, y_2) = \exp\left[- V(z_1, z_2) \right], \qquad y_1 \geq u,\quad 
    y_2 \geq u
  \end{equation}
  where $V$ is a homogeneous function of order -1,
  e.g. $V(nz_1,nz_2)=n^{-1}V(z_1,z_2)$, satisfying $V(z_1, \infty) =
  z_1^{-1}$ and $V(\infty, z_2) = z_2^{-1}$, and $z_i = - 1 / \log
  F(y_i)$, $i=1,2$.

  Contrary to the univariate case, there is no finite parametrization
  for the $V$ functions. Thus, it is common to use specific parametric
  families for $V$ such as the logistic \citep{Gumbel1960b}, the
  asymmetric logistic \citep{Tawn1988}, the negative logistic
  \citep{Galambos1975} or the asymmetric negative logistic
  \citep{Joe1990} models. Some details for these parametrisations are
  reported in Annex~\ref{sec:paramExtDep}. These models, as all models
  of the form~\eqref{eq:jointDist} are asymptotically dependent, that
  is \citep{Coles1999}

  \begin{eqnarray}
  \label{eq:asDepChi}
  \chi = \lim_{\omega\rightarrow 1} \chi(\omega) &=
  \lim_{\omega\rightarrow 1} \Pr\left[ F\left(Y_2\right)
    > \omega | F\left(Y_1\right) > \omega \right]&> 0\\
  \label{eq:asDepChiBar}
  \overline{\chi} =  \lim_{\omega\rightarrow 1}
  \overline{\chi}(\omega) &= \lim_{\omega\rightarrow 1}
  \frac{2 \log \left( 1 - \omega \right)}{\log \Pr\left[
      F\left(Y_1\right) > \omega, F\left(Y_2\right) > \omega
     \right]} -1 &= 1
\end{eqnarray}

  Other parametric families exist to consider simultaneously
  asymptotically dependent and independent cases
  \citep{Bortot1998}. However, apart from a few particular cases (see
  Section~\ref{sec:ExtValDepTest}), the data analyzed here seem to
  belong to the asymptotically dependent class.  Consequently, in this
  work, only asymptotically dependent models are considered - i.e. of
  the form~\eqref{eq:likMC}--\eqref{eq:jointDist}.

  \subsection{Inference}
  \label{subsec:inf}

  The Markov chain model is fitted using maximum censored likelihood
  estimation \citep{Ledford1996}. The contribution $L_n(y_1,y_2)$ of a
  point $(y_1, y_2)$ to the numerator of equation~\eqref{eq:likMC} is
  given by:

  \begin{equation}
    \label{eq:numlik}
    L_n(y_1,y_2) = 
    \begin{cases}
      \exp \left[-V(z_1, z_2) \right] \left[ V_1(z_1, z_2) V_2(z_1, z_2)
        - V_{12}(z_1, z_2) \right] K_1 K_2, & \text{if } y_1 > u,
      y_2 >
      u\\
      \exp \left[-V(z_1, z_2) \right] V_1(z_1, z_2) K_1, & \text{if
      } y_1 > u, y_2 \leq u\\
      \exp \left[-V(z_1, z_2) \right] V_2(z_1, z_2) K_2, & \text{if
      } y_1 \leq u, y_2 > u\\
      \exp \left[-V(z_1, z_2) \right], & \text{if } y_1 \leq u, y_2
      \leq u
    \end{cases}
  \end{equation}
  where $K_j = - \lambda_j \sigma^{-1} t_j^{1+\xi} z_j^2 \exp(1/z_j)$,
  $t_j = [1 + \xi (y_j - u) / \sigma ]_+^{-1/\xi}$ and $V_j$, $V_{12}$
  are the partial derivative with respect to the component $j$ and the
  mixed partial derivative respectively. The contribution $L_d(y_j)$
  of a point $y_j$ to the denominator of equation~\eqref{eq:likMC} is
  given by:

  \begin{equation}
    \label{eq:denlik}
    L_d(y_j) = 
    \begin{cases}
      \sigma^{-1} \lambda \left[ 1 + \xi (y_j - u) / \sigma
      \right]_+^{-1/\xi - 1}, & \text{if } y_j > u,\\
      1 - \lambda, & \text{otherwise.}
    \end{cases}
  \end{equation}

  Finally, the log-likelihood is given by:

  \begin{equation}
    \label{eq:loglikMC}
    \log L(y_1, \ldots, y_n) = \sum_{i=2}^n \log L_n(y_{i-1},y_i) -
    \sum_{i=2}^{n-1} \log L_d(y_i)
  \end{equation}
  

  \subsection{Return levels}
  \label{subsec:retlev}

  Most often, the major issue of an extreme value analysis is the
  quantile estimation. As for the POT approach, return level estimates
  can be computed. However, as all exceedances are inferred, this is
  done in a different way as the dependence between successive
  observations must be taken into account. For a stationary sequence
  $Y_1,Y_2,\ldots,Y_n$ with a marginal distribution function $F$,
  \citet{Lindgren1987} have shown that:
  
  \begin{equation}
    \label{eq:lindgren}
    \Pr\left[\max\left\{ Y_1,  Y_2, \ldots,  Y_n\right\}   \leq y  \right]
    \approx F(y)^{n \theta}
  \end{equation}
  where $\theta \in [0,1]$ is the extremal index and can be
  interpreted as the reciprocal of the mean cluster size
  \citep{Leadbetter1983} - i.e. $\theta = 0.5$ means that extreme
  (enough) events are expected to occur by pair. $\theta = 1$
  (resp. $\theta \rightarrow 0$) corresponds to the independent
  (resp. perfect dependent) case.

  As a consequence, the quantile $Q_T$ corresponding to the $T$-year
  return period is obtained by equating equation~\eqref{eq:lindgren}
  to $1-1/T$ and solving for $T$. By definition, $Q_T$ is the
  observation that is expected to be exceeded once every $T$ years,
  i.e,

  \begin{equation}
    \label{eq:rlmc}
    Q_T = u - \sigma\xi^{-1} \left( 1 - \left\{  \lambda^{-1} \left[ 1 -
        (1 - 1/T)^{1/(n\theta)} \right] \right\} ^{-\xi} \right)
  \end{equation}

  It is worth emphasizing equation~\eqref{eq:lindgren} as it has a
  large impact on both theoretical and practical aspects. Indeed, for
  the AM approach, equation~\eqref{eq:lindgren} is replaced by

  \begin{equation}
    \label{eq:bm}
    \Pr\left[\max\left\{ Y_1,  Y_2, \ldots,  Y_n\right\}   \leq y  \right]
    \approx G(y)
  \end{equation}
  where $G$ is the distribution function of the random variable $M_n =
  \max\left\{ Y_1, Y_2, \ldots, Y_n\right\}$, that is a generalized
  extreme value distribution. In particular, the
  equations~\eqref{eq:lindgren} and~\eqref{eq:bm} differ as the first
  one is fitted to the whole observations $Y_i$, while the latter is
  fitted to the AM ones. By definition, the number $n_Y$ of the $Y_i$
  observations is much larger than the size $n_M$ of the AM data
  set. Especially, for daily data, $n_Y = 365 n_M$.

  \begin{figure}
    \centering
    \includegraphics[width=0.5\textwidth]{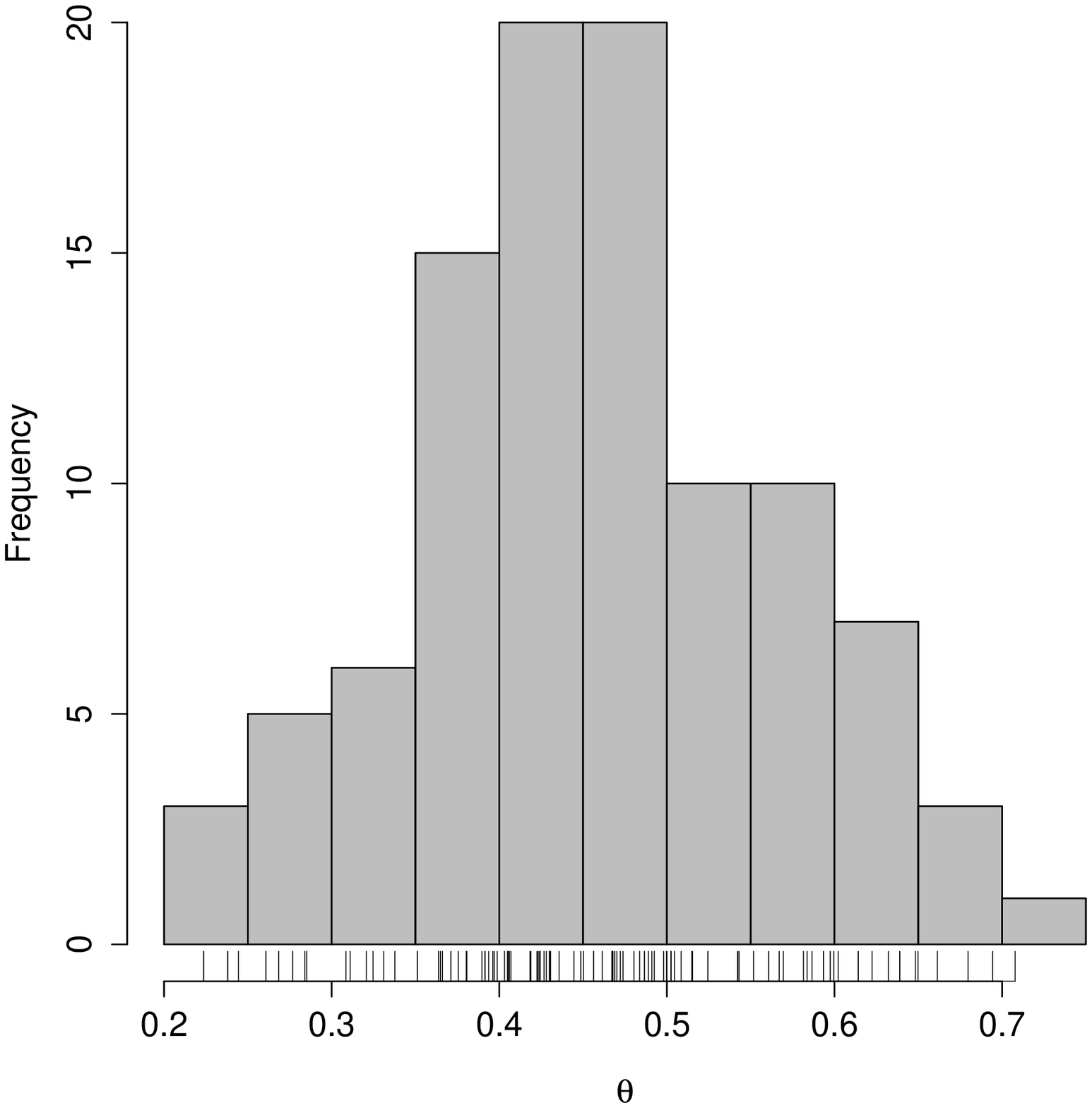}
    \caption{Histogram of the extremal index estimations from the 100
      simulated Markov Chains of length 2000.}
    \label{fig:histExi}
\end{figure}

From equation~\eqref{eq:rlmc}, the extremal index $\theta$ must be
known to obtain quantile estimates. The methodology applied in this
study is similar to the one suggested by \citet{Fawcett2006}. Once the
Markovian model is fitted, 100 Markov chains of length 2000 were
generated. For each chain, the extremal index is estimated using the
estimator proposed by \citet{Ferro2003} to avoid issues related to the
choice of declustering parameter. In particular, the
extremal index $\theta$ is estimated using the following equations:

  \begin{equation}
    \label{eq:ferro2003}
    \hat{\theta}(u) =
    \begin{cases}
      \max \left(1, \frac{2 \left[\sum_{i=1}^{N-1} \left(T_i -
              1\right) \right]^2}{\left(N - 1\right) \sum_{i=1}^{N-1}
          T_i^2} \right), &\text{if } \max\left\{ T_i: 1 \leq i \leq
      N-1 \right\} \leq 2\\
      \max\left(1, \frac{2 \left(\sum_{i=1}^{N-1} T_i \right)^2}{\left(N -
          1\right) \sum_{i=1}^{N-1} \left(T_i-1\right) \left(T_i -
          2\right)} \right), &\text{otherwise}
  \end{cases}
\end{equation}
  where $N$ is the number of observations exceeding the threshold $u$,
  $T_i$ is the inter-exceedance time, e.g. $T_i = S_{i+1} - S_i$ and
  the $S_i$ is the $i$-th exceedance time.

  Lastly, the extremal index related to a fitted Markov chain model is
  estimated using the sample mean of the 100 extremal index
  estimations.  Figure~\ref{fig:histExi} represents the histogram of
  these 100 extremal index estimations. In this study, as lots of time
  series are involved, the number and length of the simulated Markov
  chains may be too small to lead to the most accurate extremal index
  estimations; but avoid intractable CPU times. If less sites are
  considered, it is preferable to increase these two values.

  A preliminary study (not shown here) demonstrates that, for quantile
  estimation, this procedure was more accurate than estimating
  $\theta$ using the estimator of \citep{Leadbetter1983}. This
  confirms the conclusions drawn by \citet{Fawcett2005} for the
  extreme wind speed data.

  \section{Extreme Value Dependence Structure Assessment}
  \label{sec:ExtValDepTest}

  Prior to performing any estimations, it is necessary to test
  whether: (a) the first order Markov chain assumption and (b) the
  extreme value dependence structure (equation~\eqref{eq:jointDist})
  are appropriate to model successive observations above the threshold
  $u$.

     \begin{figure}
     \centering
     \includegraphics[width=\textwidth]{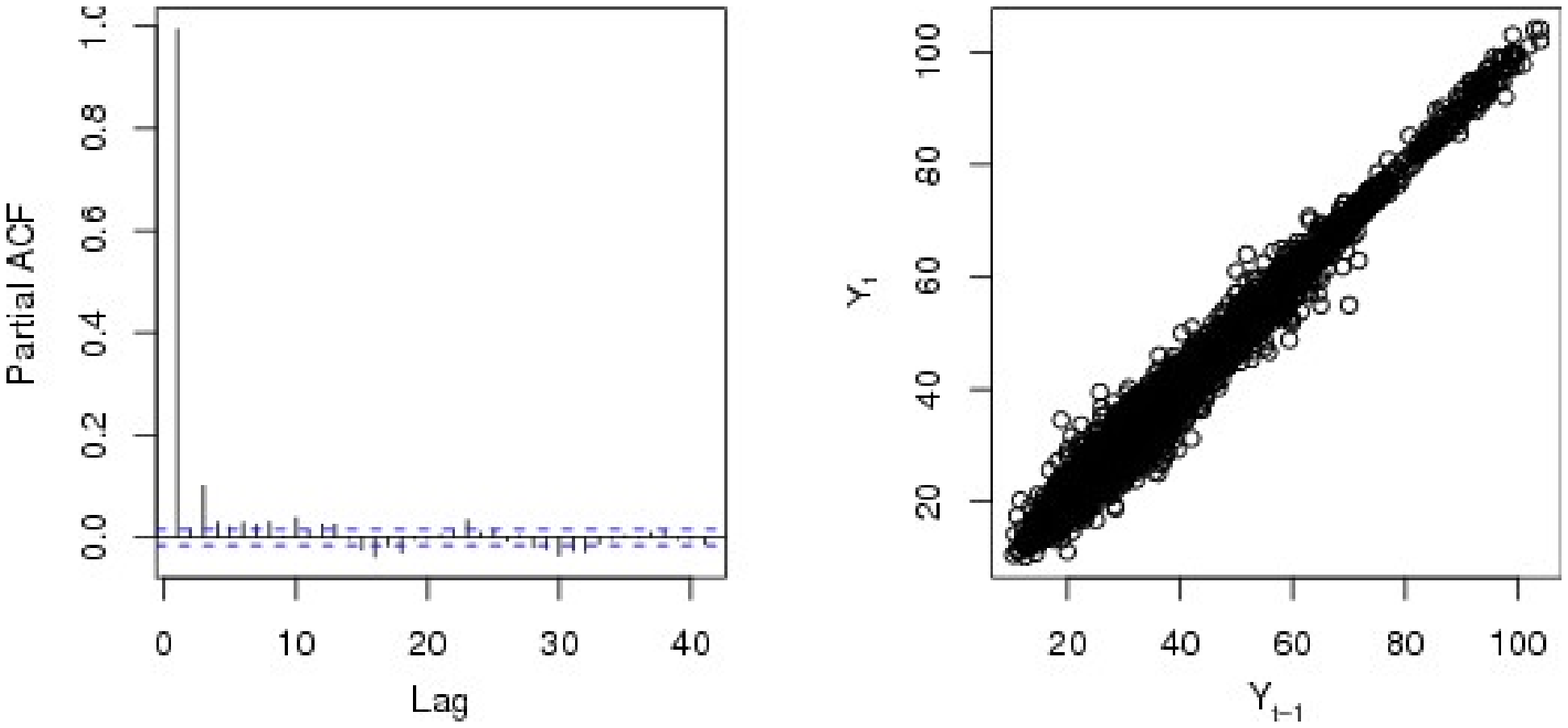}
     \caption{Autocorrelation plot (left panel) and scatterplot of the
       time series at lag 1 (right panel) for the Somme river at
       Abbeville (E6470910).}
     \label{fig:justDep1}
\end{figure}

   \begin{figure}
     \centering
     \includegraphics[width=\textwidth]{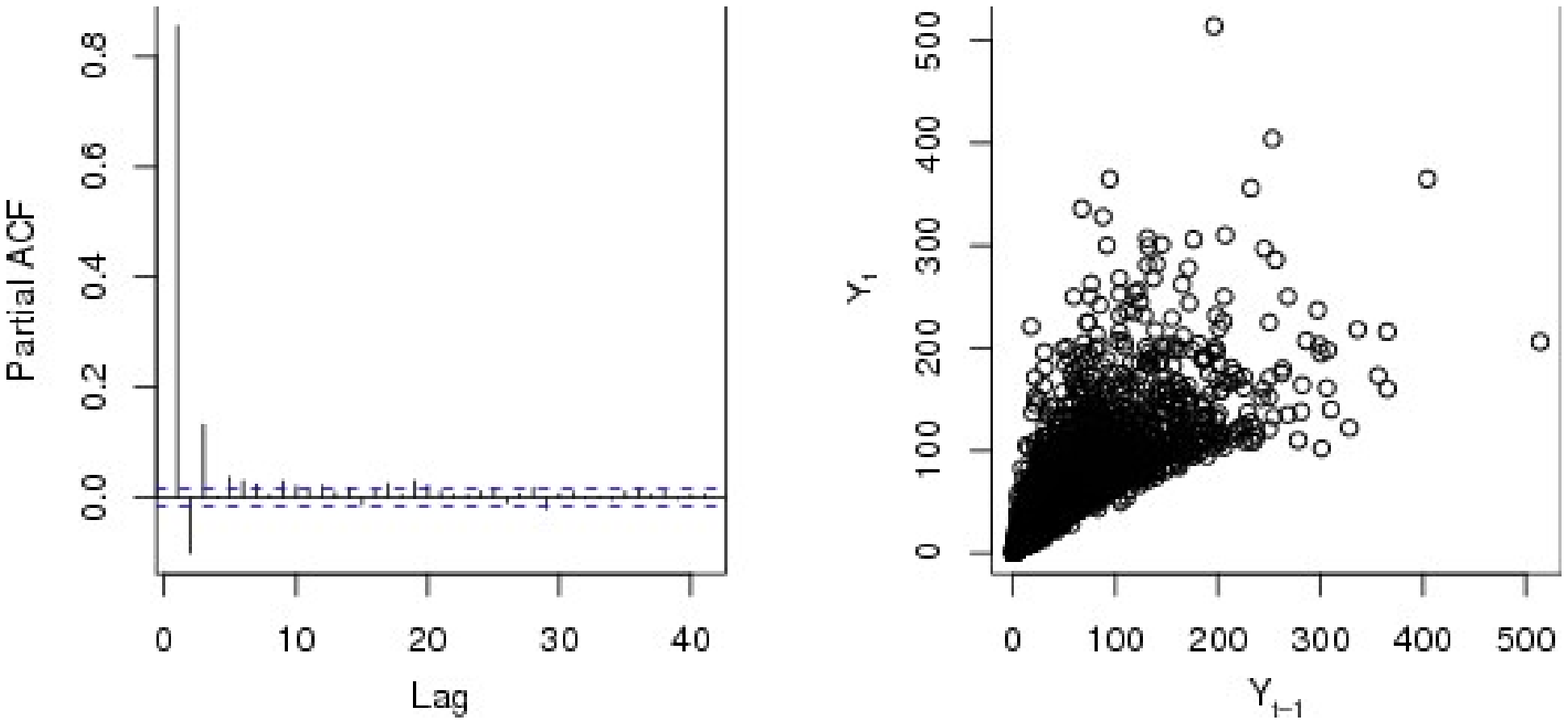}
     \caption{Autocorrelation plot (left panel) and scatterplot of the
       time series at lag 1 (right panel) for the Moselle river at
       Noirgueux (A4200630).}
     \label{fig:justDep2}
\end{figure}

Figures~\ref{fig:justDep1} and~\ref{fig:justDep2} plot the
auto-correlation functions and the scatter plots between two
consecutive observations for two different gaging stations. As the
partial autocorrelation coefficient at lag 1 is large,
Figure~\ref{fig:justDep1} and~\ref{fig:justDep2} (left panels)
corroborate the (a) hypothesis. However, as some partial
auto-correlation coefficients are significant beyond lag 1, it may
suggest that a higher-order model may be more appropriate but does not
necessarily mean that a first-order assumption is completely
flawed. Simplex plots \citep{Coles1991} (not shown) can be used to
assess the suitability of a second-order assumption over a first-order
one. For our application, it seems that a first-order model seems to
be valid - except for the five slowest dynamic catchments.

Though it is an important stage because of its consequences on
quantile estimates \citep{Ledford1996,Bortot2000}, verifying the (b)
hypothesis is a considerable task. An overwhelming dependence between
consecutive observations at finite levels is not sufficient as it does
not give any information about the dependence relation at asymptotic
levels. For instance, the overwhelming dependence at lag 1
(Figure~\ref{fig:justDep1} and~\ref{fig:justDep2}, right panels) does
certainly not justify the use of an asymptotic dependent model.

   \begin{figure}
     \centering
     \includegraphics[width=\textwidth]{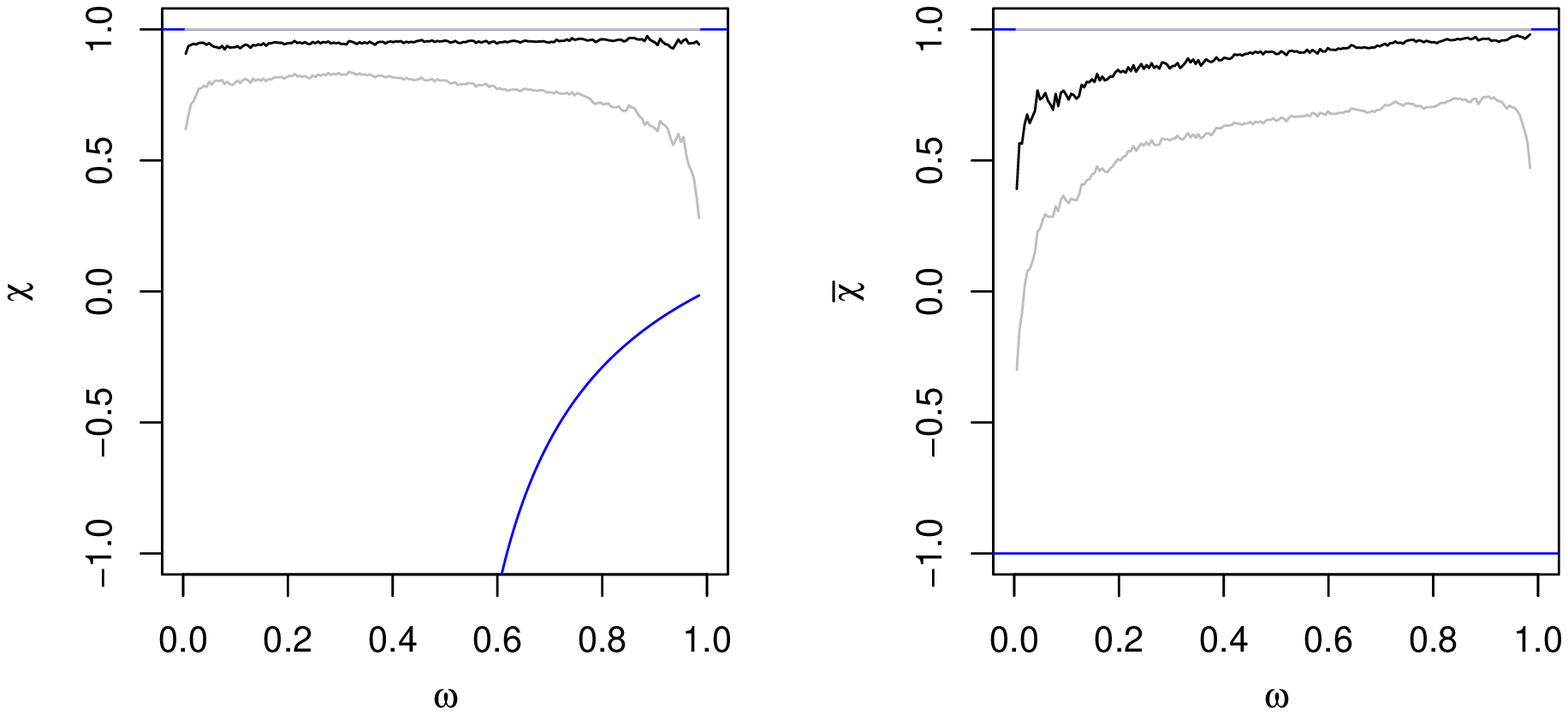}
     \caption{Plot of the $\chi$ and $\overline{\chi}$ statistics and the
       related 95\% confidence intervals for the Somme river at Abbeville
       (E6470910).  The solid blue lines are the theoretical
       bounds.}
     \label{fig:chiplot1}
\end{figure}

   \begin{figure}
     \centering
     \includegraphics[width=\textwidth]{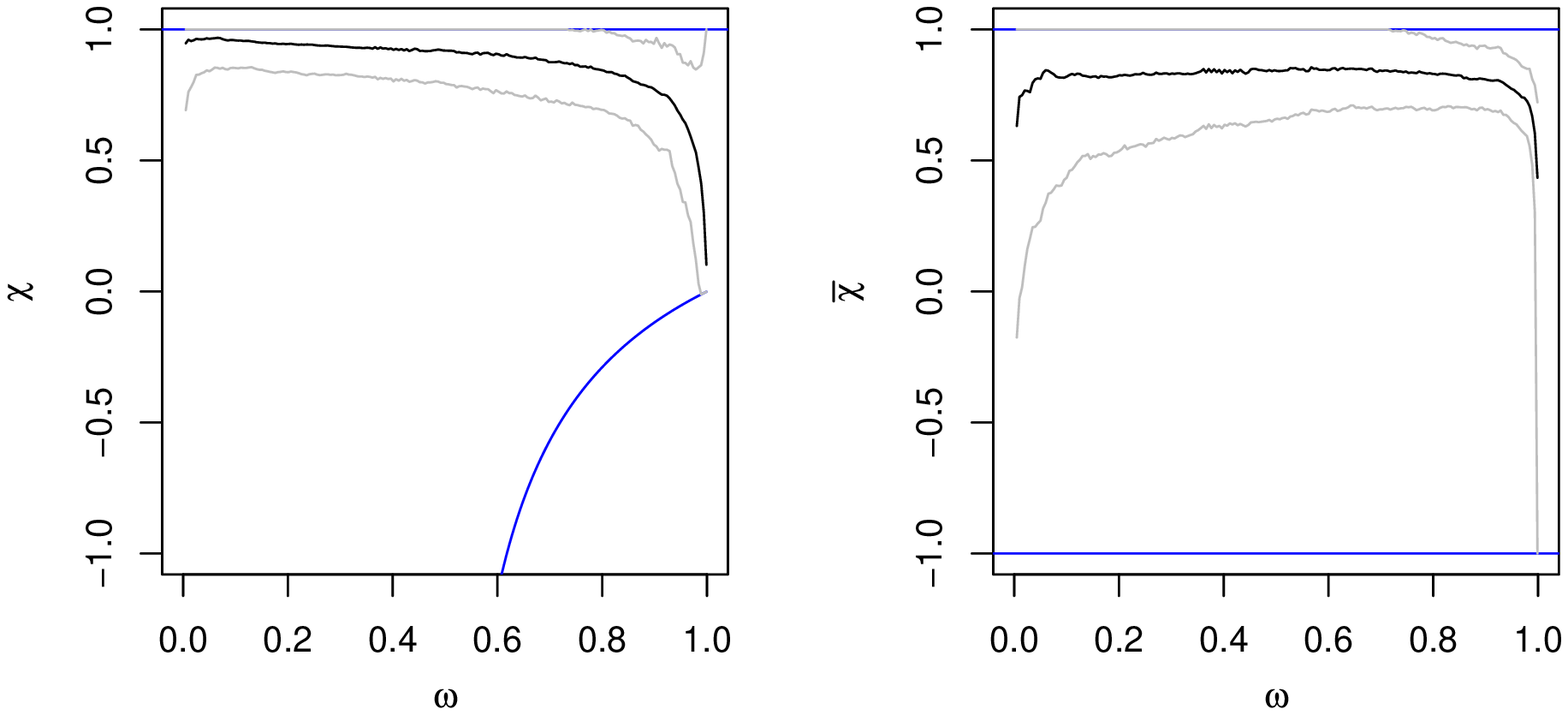}
     \caption{Plot of the $\chi$ and $\overline{\chi}$ statistics and the
       related 95\% intervals for the Moselle river at Noirgueux
       (A4200630). The solid blue lines are the theoretical
       bounds.}
     \label{fig:chiplot2}
\end{figure}
   
Figures~\ref{fig:chiplot1} and~\ref{fig:chiplot2} plot the evolution
of the $\chi(\omega)$ and $\overline{\chi}(\omega)$ statistics as
$\omega$ increases for two different sites. For these figures, the
confidence intervals are derived by bootstrapping contiguous blocks to
take into account the successive observations dependence
\citep{Ledford2003}. The $\chi(\omega)$ and $\overline{\chi}(\omega)$
statistics seem to depict two different asymptotic extremal
dependence.  From Figure~\ref{fig:chiplot1}, it seems that $\lim
\chi(\omega) \gg 0$ and $\lim \overline{\chi}(\omega) = 1$ for $\omega
\rightarrow 1$.  On the contrary, Figure~\ref{fig:chiplot2} advocates
for $\lim \chi(\omega) = 0$ and $\lim \overline{\chi}(\omega) < 1$ for
$\omega \rightarrow 1$. Consequently, Figure~\ref{fig:chiplot1} seems
to conclude for an asymptotic dependent case while
Figure~\ref{fig:chiplot2} for an asymptotic independent case.

In theory, asymptotic (in)dependence should not be assessed using
scatterplots. However, these two different features can be deduced
from Figures~\ref{fig:justDep1} and~\ref{fig:justDep2}. For
Figure~\ref{fig:justDep1}, the scatterplot $(Y_{t-1}, Y_t)$ is
increasingly less spread as the observations becomes larger; while
increasingly more spread for Figure~\ref{fig:justDep2}. In other
words, for the first case, the dependence seems to become stronger at
larger levels while this is the contrary for the second case.

   \begin{table}
     \caption{$\chi(\omega)$ statistics for all
       stations. $\omega=0.98, 0.985, 0.99$.}
     \label{tab:chiStats}
     \centering
     \begin{tabular}{rrrrrrrrrr}
       \hline
       \multirow{2}*{Stations} & \multicolumn{2}{c}{$\omega = 0.98$} &&
       \multicolumn{2}{c}{$\omega = 0.985$} && \multicolumn{2}{c}{$\omega = 0.99$}\\
       \cline{2-3} \cline{5-6} \cline{8-9}
       & $\chi(\omega)$ & 95\% C.I. && $\chi(\omega)$ & 95\% C.I. && $\chi(\omega)$ &
       95\% C.I.\\
       \hline
       A3472010 & 0.67 & (-0.02, 1.00) && 0.60 & (-0.02, 1.00) && 0.57 & (-0.01, 1.00)\\
       A4200630 & 0.53 & ( 0.21, 0.81) && 0.45 & ( 0.07, 0.77) && 0.38 & (-0.01, 0.76)\\
       A4250640 & 0.55 & ( 0.27, 0.82) && 0.49 & ( 0.18, 0.76) && 0.41 & ( 0.02, 0.71)\\
       A5431010 & 0.44 & (-0.02, 1.00) && 0.44 & (-0.02, 1.00) && 0.41 & (-0.01, 1.00)\\
       A5730610 & 0.59 & ( 0.25, 0.94) && 0.56 & ( 0.20, 0.90) && 0.50 & ( 0.07, 0.97)\\
       A6941010 & 0.62 & ( 0.22, 0.99) && 0.60 & ( 0.16, 1.00) && 0.56 & ( 0.06, 1.00)\\
       A6941015 & 0.63 & ( 0.29, 0.95) && 0.60 & ( 0.20, 0.96) && 0.58 & ( 0.17, 0.98)\\
       D0137010 & 0.39 & ( 0.04, 0.69) && 0.33 & (-0.02, 0.67) && 0.28 & (-0.01, 0.69)\\
       D0156510 & 0.59 & ( 0.25, 0.88) && 0.55 & ( 0.20, 0.86) && 0.53 & ( 0.14, 0.92)\\
       E1727510 & 0.62 & ( 0.18, 0.91) && 0.59 & ( 0.16, 0.93) && 0.47 & (-0.01, 0.89)\\
       E1766010 & 0.63 & ( 0.23, 0.98) && 0.59 & ( 0.17, 0.96) && 0.54 & ( 0.09, 0.96)\\
       E3511220 & 0.59 & ( 0.10, 1.00) && 0.53 & (-0.02, 1.00) && 0.50 & (-0.01, 0.99)\\
       E4035710 & 0.77 & ( 0.02, 1.00) && 0.68 & (-0.02, 1.00) && 0.60 & (-0.01, 1.00)\\
       E5400310 & 0.88 & ( 0.30, 1.00) && 0.89 & ( 0.29, 1.00) && 0.83 & ( 0.13, 1.00)\\
       E5505720 & 0.91 & ( 0.24, 1.00) && 0.87 & ( 0.09, 1.00) && 0.86 & ( 0.02, 1.00)\\
       E6470910 & 0.96 & ( 0.40, 1.00) && 0.94 & ( 0.25, 1.00) && 0.98 & ( 0.00, 1.00)\\
       H0400010 & 0.84 & ( 0.12, 1.00) && 0.83 & ( 0.02, 1.00) && 0.78 & (-0.01, 1.00)\\
       H1501010 & 0.82 & ( 0.36, 1.00) && 0.90 & ( 0.39, 1.00) && 0.84 & ( 0.26, 1.00)\\
       H2342010 & 0.68 & ( 0.31, 1.00) && 0.67 & ( 0.25, 1.00) && 0.60 & ( 0.11, 1.00)\\
       H5071010 & 0.75 & ( 0.30, 1.00) && 0.76 & ( 0.22, 1.00) && 0.75 & ( 0.15, 1.00)\\
       H5172010 & 0.80 & ( 0.47, 1.00) && 0.77 & ( 0.42, 1.00) && 0.73 & ( 0.30, 1.00)\\
       H6201010 & 0.69 & ( 0.29, 1.00) && 0.69 & ( 0.14, 1.00) && 0.69 & ( 0.08, 1.00)\\
       H7401010 & 0.85 & ( 0.46, 1.00) && 0.85 & ( 0.38, 1.00) && 0.81 & ( 0.27, 1.00)\\
       I9221010 & 0.67 & ( 0.23, 1.00) && 0.66 & ( 0.19, 1.00) && 0.59 & ( 0.04, 1.00)\\
       J0621610 & 0.61 & ( 0.25, 0.92) && 0.58 & ( 0.20, 0.94) && 0.51 & ( 0.08, 0.91)\\
       K0433010 & 0.59 & ( 0.22, 0.91) && 0.54 & ( 0.15, 0.89) && 0.45 & ( 0.00, 0.85)\\
       K0454010 & 0.71 & ( 0.37, 1.00) && 0.67 & ( 0.24, 1.00) && 0.65 & ( 0.14, 1.00)\\
       K0523010 & 0.62 & (-0.02, 1.00) && 0.58 & (-0.02, 1.00) && 0.53 & (-0.01, 1.00)\\
       K0550010 & 0.61 & ( 0.22, 0.94) && 0.57 & ( 0.15, 0.94) && 0.54 & ( 0.07, 1.00)\\
       K0673310 & 0.67 & ( 0.24, 1.00) && 0.65 & ( 0.18, 1.00) && 0.66 & ( 0.07, 1.00)\\
       K0910010 & 0.65 & (-0.02, 1.00) && 0.61 & (-0.02, 1.00) && 0.58 & (-0.01, 1.00)\\
       K1391810 & 0.68 & ( 0.27, 1.00) && 0.64 & ( 0.16, 0.98) && 0.60 & ( 0.06, 0.96)\\
       K1503010 & 0.69 & ( 0.38, 0.98) && 0.67 & ( 0.30, 0.98) && 0.64 & ( 0.23, 1.00)\\
       K2330810 & 0.68 & ( 0.29, 1.00) && 0.66 & ( 0.22, 1.00) && 0.62 & ( 0.09, 1.00)\\
       K2363010 & 0.65 & ( 0.26, 0.98) && 0.66 & ( 0.16, 1.00) && 0.61 & ( 0.01, 1.00)\\
       K2514010 & 0.61 & ( 0.24, 1.00) && 0.61 & ( 0.21, 1.00) && 0.58 & ( 0.12, 1.00)\\
       K2523010 & 0.53 & (-0.02, 1.00) && 0.53 & (-0.02, 1.00) && 0.51 & (-0.01, 1.00)\\
       K2654010 & 0.68 & ( 0.37, 1.00) && 0.68 & ( 0.31, 1.00) && 0.60 & ( 0.10, 1.00)\\
       K2674010 & 0.60 & ( 0.25, 0.89) && 0.58 & ( 0.22, 0.94) && 0.54 & ( 0.08, 0.95)\\
       K2871910 & 0.62 & ( 0.26, 0.95) && 0.57 & ( 0.15, 0.94) && 0.56 & ( 0.10, 0.97)\\
       K2884010 & 0.62 & ( 0.25, 1.00) && 0.57 & ( 0.17, 0.97) && 0.59 & ( 0.16, 1.00)\\
       K3222010 & 0.56 & ( 0.21, 0.90) && 0.53 & ( 0.18, 0.93) && 0.46 & ( 0.11, 0.89)\\
       K3292020 & 0.59 & ( 0.27, 0.91) && 0.57 & ( 0.17, 0.91) && 0.48 & ( 0.07, 0.90)\\
       K4470010 & 0.76 & ( 0.39, 1.00) && 0.77 & ( 0.40, 1.00) && 0.73 & ( 0.27, 1.00)\\
       K5090910 & 0.64 & ( 0.27, 0.93) && 0.64 & ( 0.26, 0.96) && 0.58 & ( 0.12, 0.98)\\
       K5183010 & 0.57 & ( 0.14, 0.91) && 0.56 & ( 0.15, 0.96) && 0.53 & ( 0.06, 0.97)\\
       K5200910 & 0.63 & ( 0.24, 0.93) && 0.62 & ( 0.20, 0.95) && 0.56 & ( 0.11, 0.97)\\
       L0140610 & 0.73 & ( 0.23, 1.00) && 0.66 & ( 0.15, 1.00) && 0.58 & (-0.01, 1.00)\\
       L0231510 & 0.59 & ( 0.16, 0.91) && 0.55 & ( 0.11, 0.92) && 0.53 & (-0.01, 0.92)\\
       L0400610 & 0.74 & (-0.02, 1.00) && 0.65 & (-0.02, 1.00) && 0.61 & (-0.01, 1.00)\\
       \hline
     \end{tabular}
\end{table}

   \begin{figure}
     \centering
     \includegraphics[width=\textwidth]{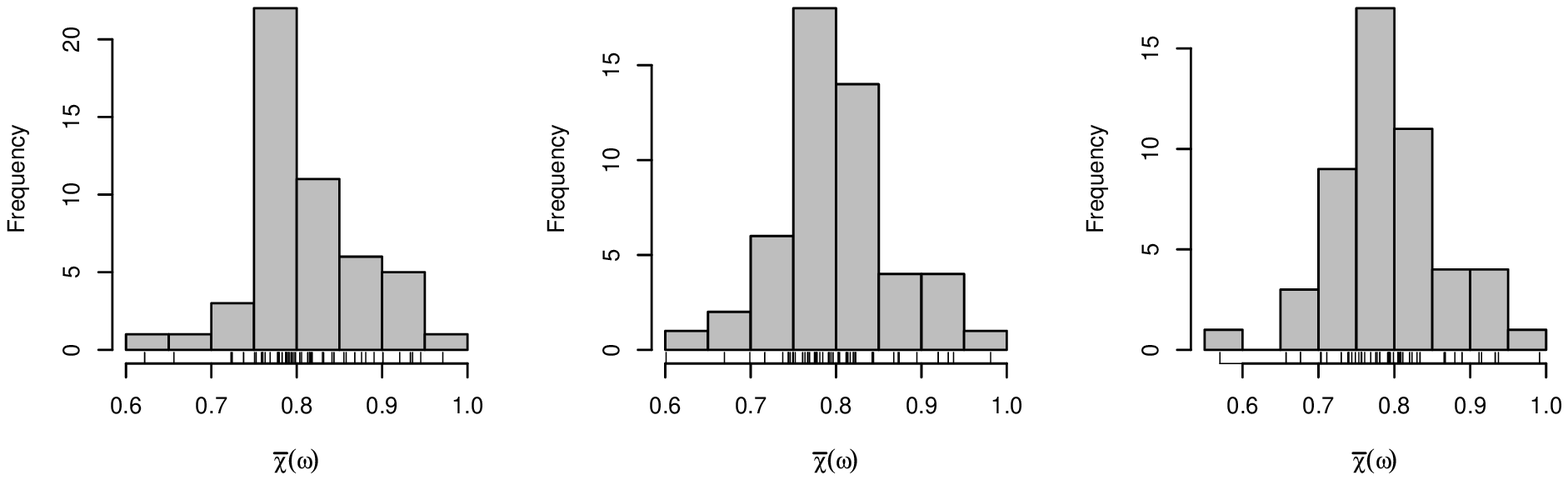}
     \caption{Histograms of the $\overline{\chi}(\omega)$ statistics
       for different $\omega$ values. Left panel: $\omega=0.98$,
       middle panel: $\omega=0.985$ and right panel: $\omega=0.99$.}
     \label{fig:histChiBar}
\end{figure}

Two specific cases for different asymptotic dependence structures were
illustrated. Table~\ref{tab:chiStats} shows the evolution of the
$\chi(\omega)$ statistics as $\omega$ increases for all the sites
under study. Most of the stations have significantly positive
$\chi(\omega)$ values. In addition, only 13 sites have a 95\%
confidence interval that contains the 0 value.  For 9 of these
stations, the 95\% confidence intervals correspond to the theoretical
lower and upper bounds; so that uncertainties are too large to
determine the extremal dependence class. For the $\overline{\chi}$
statistic, results are less clear-cut. Figure~\ref{fig:histChiBar}
represents the histograms for $\overline{\chi}(\omega)$ for successive
$\omega$ values. Despite only a few observations being close to 1,
most of the stations have a $\overline{\chi}(\omega)$ value greater
than 0.75.  These values can be considered as significantly high as
$-1 < \overline{\chi}(\omega) \leq 1$, for all $\omega$. Consequently,
models of the form~\eqref{eq:likMC}--\eqref{eq:jointDist} may be
suited to model the extremal dependence between successive
observations.

Other methods exist to test the extremal dependence but were
unconvincing for our application \citep{Ledford2003,Falk2006}. Indeed,
the approach of \citet{Falk2006} does not take into account the
dependence between $Y_{t-1}$ and $Y_t$; while the test of
\citet{Ledford2003} appears to be poorly discriminatory for our case
study.

\section{Performance of the Markovian Models on Quantile Estimation}
\label{sec:perfMark}

\subsection{Comparison between Markovian estimators}
\label{subsec:compMark}

In this section, the performance of six different extremal dependence
structures is analyzed on the 50 gaging stations introduced in
section~\ref{sec:intro}. These models are: $log$ for the logistic,
$nlog$ for the negative logistic, $mix$ for the mixed models and their
relative asymmetric counterparts - e.g. $alog$, $anlog$ and $amix$. To
assess the impact of the dependence structure on flood peak
estimation, the efficiency of each model to estimate quantiles with
return periods 2, 10, 20, 50 and 100 years is evaluated.

As practitioners often have to deal with small record lengths in
practice, the performance of the Markovian models is analyzed on all
sub time series of length 5, 10, 15 and 20 years. Finally, to assess
the efficiency for all the gaging stations considered in this study,
the normalized bias ($\mathbf{nbias}$), the variance ($\mathbf{var}$)
and the normalized mean squared error ($\mathbf{nmse}$) are computed:

   \begin{eqnarray}
     \label{eq:nbias}
     nbias &=& \frac{1}{N}\sum_{i=1}^N \frac{\hat{Q}_{i,T} - Q_T}{Q_T}\\
     \label{eq:sd}
     var &=& \frac{1}{N-1} \sum_{i=1}^N \left( \frac{\hat{Q}_{i,T} - Q_T}{Q_T} - nbias \right)^2\\
     \label{eq:nmse}
     nmse &=& \frac{1}{N}\sum_{i=1}^N \left(\frac{\hat{Q}_{i,T} - Q_T}{Q_T}
     \right)^2
\end{eqnarray}
   where $Q_T$ is the benchmark $T$-year return level and
   $\hat{Q}_{i,T}$ is the $i$-th estimate of $Q_T$.

   \begin{figure}
     \centering
     \includegraphics[width=\textwidth]{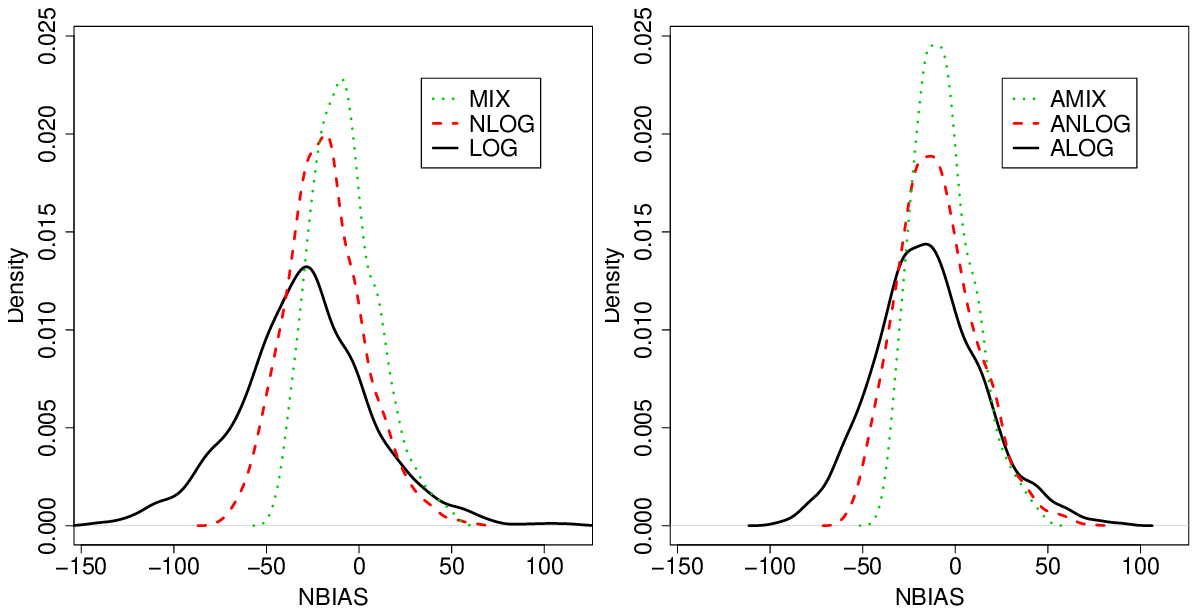}
     \caption{Densities of the normalized biases of $Q_{20}$ estimates for
       the symmetric Markovian models (left panel) and the asymmetric
       ones (right panel). Target site record length: 5 years.}
     \label{fig:Q20Mark5years}
\end{figure}

Figure~\ref{fig:Q20Mark5years} depicts the $nbias$ densities for
$Q_{20}$ with a record length of 5 years. It is overwhelming that the
extremal dependence structure has a great impact on the estimation of
$Q_{20}$. Comparing the two panels, it can be noticed that the
symmetric dependence structures give spreader densities; that is, more
variable estimates. Independently of the symmetry,
Figure~\ref{fig:Q20Mark5years} shows that the mixed dependence family
is more accurate.

   \begin{table}
     \caption{Several characteristics of the Markovian estimators on
       $Q_{50}$ estimation as the record length increases. Standard errors are reported in brackets.}
     \label{tab:sumTabMarkQ50}
     \centering
     \begin{tabular}{lcccccccccccccccc}
       \hline
       \multirow{2}*{Model} & \multicolumn{3}{c}{5 years} &&
       \multicolumn{3}{c}{10 years} && \multicolumn{3}{c}{15 years} &&
       \multicolumn{3}{c}{20 years}\\
       \cline{2-4} \cline{6-8} \cline{10-12} \cline{14-16}
       & $nbias$ & $var$ & $nmse$ && $nbias$ & $var$ & $nmse$ && $nbias$ &
       $var$ & $nmse$ &&  $nbias$ & $var$ & $nmse$\\
       \hline
       $log$   & -0.35  & 0.54    & 0.66    && -0.32   & 0.32   & 0.42    && -0.30  & 0.23   & 0.32   && -0.28  & 0.17   & 0.25 \\
               & (16e-3)& (22e-3) & (18e-3) &&  (12e-3)& (12e-3)& (14e-3) &&(11e-3) & (9e-3) &(12e-3) && (9e-3) & (7e-3) &  (11e-3)\\
       $nlog$  & -0.21  & 0.20    & 0.24    && -0.20   & 0.11   & 0.15    && -0.18  & 0.08   & 0.12   && -0.18  & 0.06   & 0.09 \\
               & (10e-3)& (7e-3)  & (11e-3) && (7e-3)  & (4e-3) & (9e-3)  && (6e-3) & (3e-3) & (8e-3) && (5e-3) & (2e-3) & (7e-3)\\
       $mix$   & -0.08  & 0.14    & 0.14    && -0.07   & 0.08   & 0.08    && -0.06  & 0.05   & 0.06   && -0.05  & 0.04   & 0.04 \\
               & (8e-3) & (5e-3)  & (8e-3)  && (6e-3)  & (2e-3) & (6e-3)  && (5e-3) & (2e-3) & (5e-3) && (4e-3) & (1e-3) & (5e-3)\\
       $alog$  & -0.15  & 0.39    & 0.41    && -0.13   & 0.22   & 0.24    && -0.11  & 0.16   & 0.17   && -0.10  & 0.12   & 0.13 \\
               & (14e-3)& (15e-3) & (14e-3) && (10e-3) & (9e-3) & (11e-3) && (9e-3) & (6e-3) & (9e-3) && (8e-3) & (4e-3) & (8e-3)\\
       $anlog$ & -0.10  & 0.20    & 0.21    && -0.09   & 0.11   & 0.12    && -0.08  & 0.08   & 0.09   && -0.08  & 0.06   & 0.06 \\
               & (10e-3)& (7e-3)  & (10e-3) && (7e-3)  & (4e-3) & (8e-3)  && (6e-3) & (2e-3) & (6e-3) && (5e-3) & (2e-3) & (6e-3)\\
       $amix$  & -0.06  & 0.11    & 0.12    && -0.05   & 0.06   & 0.06    && -0.04  & 0.04   & 0.05   && -0.03  & 0.03   & 0.03 \\
               & (7e-3) & (4e-3)  & (7e-3)  && (6e-3)  & (2e-3) & (6e-3)  && (5e-3) & (1e-3) & (5e-3) && (4e-3) & (1e-3) & (4e-3)\\
       \hline
     \end{tabular}
\end{table}

Table~\ref{tab:sumTabMarkQ50} shows the $nbias$, $var$ and $nmse$
statistics for all the Markovian estimators as the record length
increases for quantile $Q_{50}$. This table confirms results derived
from Figure~\ref{fig:Q20Mark5years}. Indeed, the asymmetric dependence
structures give less variable and biased estimates - as their $nbias$
and $var$ statistics are smaller. In addition, whatever the record
length is, the Markovian models perform with the same hierarchy; that
is the $mix$ and $amix$ models are by far the most accurate
estimators - i.e.\ with the smallest $nmse$ values. The same results
(not shown) have been found for other quantiles.

From an hydrological point of view, these two results are not
surprising. The symmetric models suppose that the variables $Y_t$ and
$Y_{t+1}$ are exchangeable. In our context, exchangeability means that
the time series are reversible - e.g.\ the time vector direction has
no importance. When dealing with AM or POT and stationary time series,
it is a reasonable hypothesis. For example, the MLE remains the same
with any permutations of the AM/POT sample. However, when modeling all
exceedances, the time direction can not be considered as reversible as
flood hydrographs are clearly non symmetric.




   \begin{figure}
     \centering
     \includegraphics[width=\textwidth]{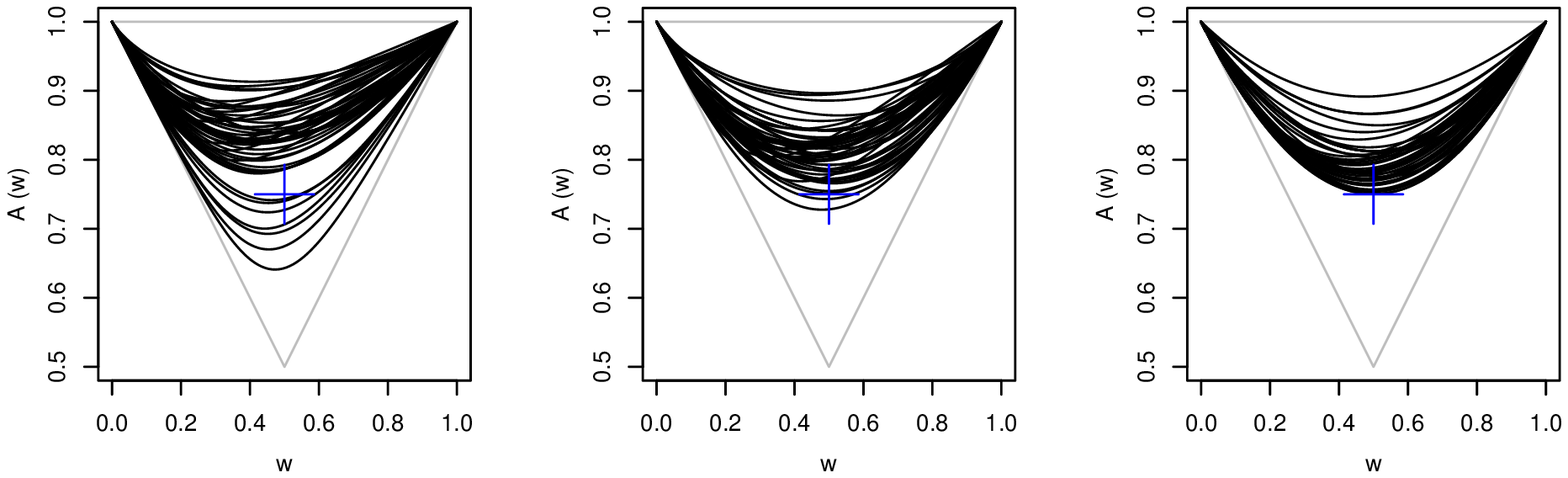}
     \caption{Representation of the Pickands' dependence functions for
       the 50 gaging stations. Left panel : $alog$, middle panel:
       $anlog$ and right panel: $amix$. ``+'' represents the theoretical
       dependence bound for the $amix$ model.}
     \label{fig:pickands}
\end{figure}


The Pickands' dependence function $A(\omega)$ \citep{Pickands1981} is
another representation for the extremal dependence structure for any
extreme value distribution. $A(\omega)$ is related to the $V$ function
in equation~\eqref{eq:jointDist} as follows:

\begin{equation}
  \label{eq:pickDep}
  A(\omega) = \frac{V(z_1,z_2)}{z_1^{-1}+z_2^{-1}}, \qquad \omega =
  \frac{z_1}{z_1+z_2}
\end{equation}

Figure~\ref{fig:pickands} represents the Pickands' dependence function
for all the gaging stations and the three asymmetric Markovian
models. One major specificity of the mixed models is that these models
can not account for perfect dependence cases. In particular, the
Pickands' dependence functions for the mixed models satisfy $A(0.5)
\geq 0.75$ while $A(0.5) \in [0.5,1]$ for the logistic and negative
logistic models. From Figure~\ref{fig:pickands}, it can be seen that
only few stations have a dependence function that could not be modeled
by the $amix$ model. Therefore, the dependence range limitation of the
$amix$ model does not seem too restrictive.


In this section, the effect of the extremal dependence structure has
been assessed. It has been established that the symmetric models are
hydrologically inconsistent as they could not reproduce the flood
event asymmetry. In addition, for all the quantiles analyzed, the
asymmetric mixed model is the most accurate for flood peak
estimations. Therefore, in the remainder of this section, only the
$amix$ model will be compared to conventional POT estimators.

\subsection{Comparison between $amix$ and conventional POT estimators}
\label{subsec:compAmixClass}

In this section, the performance of the $amix$ estimator is compared
to the estimators usually used in flood frequency analysis. For this
purpose, the quantile estimates derived from the Maximum Likelihood
Estimator ($\mathbf{MLE}$), the Unbiased and Biased Probability
Weighted moments estimators \citep{Hosking1987} ($\mathbf{PWU}$ and
$\mathbf{PWB}$ respectively) are considered.

   \begin{figure}
     \centering
     \includegraphics[width=\textwidth]{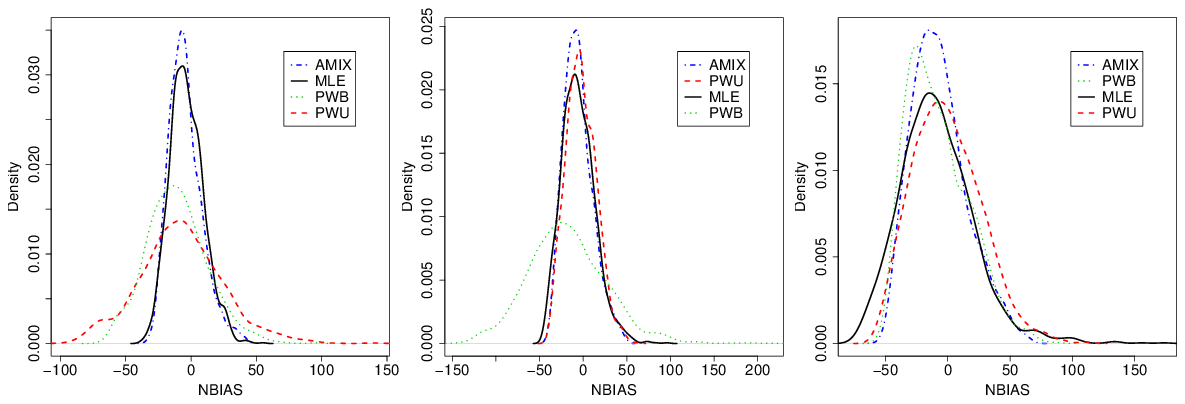}
     \caption{Densities of the normalized biases for the $amix$ model and
       the $MLE$, $PWU$, and $PWB$ estimators for quantiles $Q_5$ (left
       panel), $Q_{10}$ (middle panel) and $Q_{20}$ (right panel). Record
       length: 5 years.}
     \label{fig:amixClass}
\end{figure}

Figure~\ref{fig:amixClass} depicts the $nbias$ densities for the
$amix$, $MLE$, $PWU$ and $PWB$ estimators related to the $Q_5$,
$Q_{10}$ and $Q_{20}$ estimations with a record length of 5 years. It
can be seen that $amix$ is the most accurate model for all
quantiles. Indeed, the $amix$ $nbias$ densities are the most sharp
with a mode close to 0. Focusing only on ``classical'' estimators
(e.g. $MLE$, $PWU$ and $PWB$), there is no estimator that perform
better than any other anytime. These two results advocate the use of
the $amix$ model.

   \begin{table}
     \caption{Several characteristics of the $amix$, $MLE$, $PWU$ and
       $PWB$ estimators for $Q_{50}$ estimation as the record length
       increases. Standard errors are reported in brackets.}
     \label{tab:amixClassQ50}
     \centering
     \begin{tabular}{lccccccccccccccc}
       \hline
       \multirow{2}*{Model} & \multicolumn{3}{c}{5 years} &&
       \multicolumn{3}{c}{10 years} && \multicolumn{3}{c}{15 years} &&
       \multicolumn{3}{c}{20 years}\\
       \cline{2-4} \cline{6-8} \cline{10-12} \cline{14-16}
       & $nbias$ & $var$ & $nmse$ && $nbias$ & $var$ & $nmse$ && $nbias$ &
       $var$ & $nmse$ &&  $nbias$ & $var$ & $nmse$\\
       \hline
       $amix$ & -0.06  & 0.11   & 0.12  && -0.05  & 0.06   & 0.07   && -0.04 & 0.04   & 0.05   && -0.04 & 0.03   & 0.03 \\
              & (8e-3) & (4e-3) & (8e-3)&& (6e-3) & (2e-3) & (6e-3) &&(5e-3) & (1e-3) & (5e-3) &&(4e-3) & (1e-3) & (4e-3) \\
       $MLE$  & -0.13  & 0.25   & 0.27  && -0.14  & 0.13   & 0.14   && -0.13 & 0.08   & 0.10   && -0.11 & 0.05   & 0.07 \\
              & (12e-3)& (15e-3)&(12e-3)&& (8e-3) & (6e-3) & (9e-3) &&(7e-3) & (3e-3) & (7e-3) &&(5e-3) & (2e-3) & (6e-3) \\
       $PWU$  &  0.08  & 0.30   & 0.31  && -0.01  & 0.15   & 0.15   && -0.03 & 0.10   & 0.10   && -0.03 & 0.06   & 0.06 \\
              & (13e-3)& (13e-3)&(13e-3)&& (9e-3) & (6e-3) & (9e-3) &&(7e-3) & (3e-3) & (7e-3) &&(6e-3) & (2e-3) & (6e-3) \\
       $PWB$  & -0.07  & 0.20   & 0.21  && -0.10  & 0.11   & 0.12   && -0.11 & 0.08   & 0.09   && -0.10 & 0.05   & 0.06 \\
              & (10e-3)& (8e-3) &(11e-3)&& (7e-3) & (4e-3) & (8e-3) &&(6e-3) & (2e-3) & (7e-3) &&(5e-3) & (1e-3) & (6e-3) \\
       \hline
     \end{tabular}
\end{table}

Table~\ref{tab:amixClassQ50} shows the performance of each estimator
to estimate $Q_{50}$ as the record length increases. It can be seen
that the $amix$ model performs better than the conventional estimators
for the whole range of record lengths analyzed. First, $amix$ has the
same bias than the conventional estimators. Thus, the $amix$
dependence structure seems to be suited to estimate flood quantile
estimates. Second, because of its smaller variance, $amix$ is more
accurate than $MLE$, $PWU$ and $PWB$ estimators. This smaller variance
is mainly a result of all of the exceedances (not only cluster maxima)
being used in the inference procedure. Consequently, the $amix$ model
has a smaller $nmse$ - around half of the conventional models ones.

   \begin{figure}
     \centering
     \includegraphics[angle=-90,width=\textwidth]{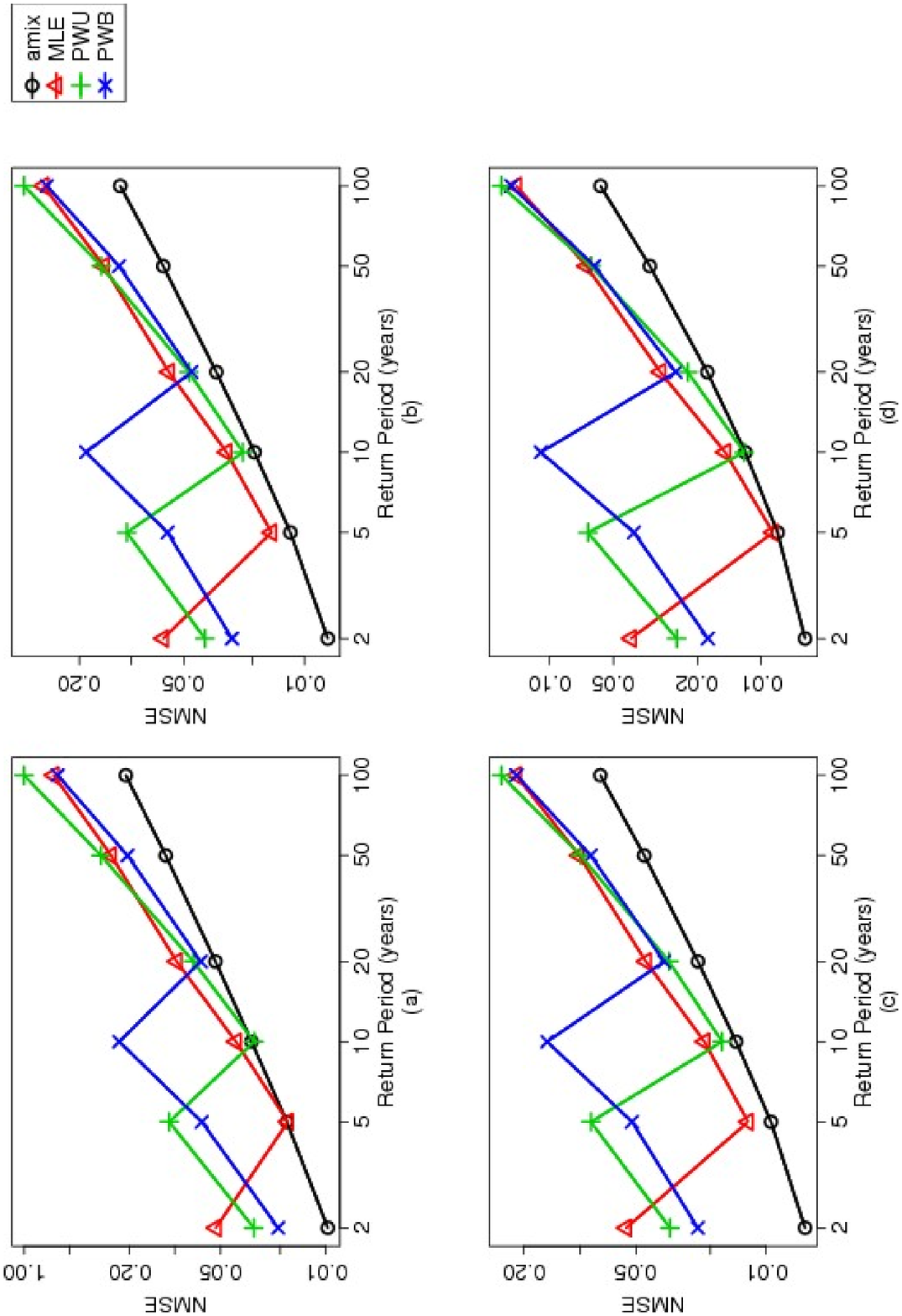}
     \caption{Evolution of the $nmse$ as the return period increases for
       the $amix$, $MLE$, $PWU$ and $PWB$ estimators. Record length: (a)
       5 years, (b) 10 years, (c) 15 years and (d) 20 years.}
     \label{fig:nmseVsRetPer}
\end{figure}

Figure~\ref{fig:nmseVsRetPer} shows the evolution of the $nmse$ as the
return period increases for the $amix$, $MLE$, $PWU$ and $PWB$ models.
This figure corroborates the conclusions drawn from
Figure~\ref{fig:amixClass} and Table~\ref{tab:amixClassQ50}. It can be
seen that the $amix$ model has the smallest $nmse$, independently of
the return period and the record length. In addition, the $amix$
becomes increasingly more efficient as the return period increases -
mostly for return periods greater than 20 years. While the
conventional estimators present an erratic $nmse$ behavior as the
return period increases, the $amix$ model is the only one that has a
smooth evolution. To conclude, these results confirm that the $amix$
model clearly improves flood peak quantile estimates - especially for
large return periods.

\section{Inference on Other Flood Characteristics}
\label{sec:infFlood}

As all exceedances are modeled using a first order Markov chain, it is
possible to infer other quantities than flood peaks - e.g.\ volume and
duration. In this section, the ability of these Markovian models to
reproduce the flood duration is analyzed.  For this purpose, the most
severe flood hydrographs within each year are considered and
normalized by their peak values. Consequently, from this observed
normalized hydrograph set, two flood characteristics derived from a
data set of hydrographs \citep{FEH1999,Sauquet2008} are considered:
(a) the duration $d_{mean}$ above 0.5 of the normalized hydrograph set
mean and (b) the median $d_{med}$ of the durations above 0.5 of each
normalized hydrograph.

\subsection{Global Performance}
\label{subsec:globPerf}

   \begin{figure}
     \centering
     \includegraphics[angle=-90,width=\textwidth]{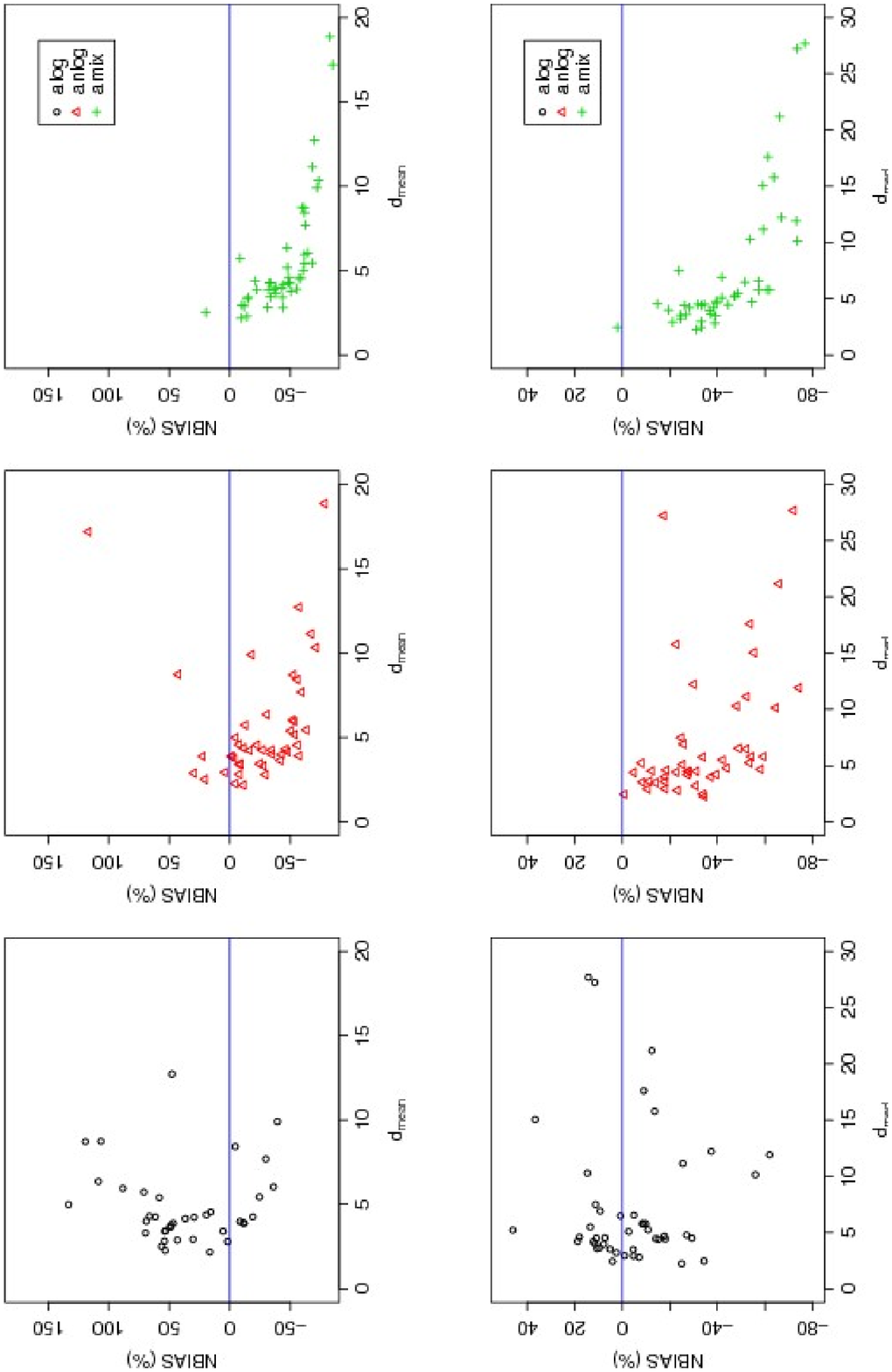}
     \caption{$d_{mean}$ and $d_{med}$ normalized biases in function of
       the theoretical values for the three asymmetric Markovian
       models.}
     \label{fig:durBias}
\end{figure}

Figure~\ref{fig:durBias} plots the flood durations $d_{mean}$ and
$d_{med}$ biases derived from the three asymmetric Markovian models in
function of their empirical estimates. It can be seen that no model
leads to accurate flood duration estimations. In addition, the extremal
dependence structure has a clear impact on these estimations. In
particular, the $anlog$ and $amix$ models seem to underestimate the
flood durations, while the $alog$ model leads to overestimations.
Consequently, two different conclusions can be drawn. First, as large
durations are poorly estimated, higher order Markov chains may be of
interest. However, this is a considerable task as higher dimensional
multivariate extreme value distributions often lead to numerical
problems. Instead of considering higher order, another alternative may
be to change daily observations for $d$-day observations - where $d$
is larger than 1. Second, it is overwhelming that the extremal
dependence structure affects the flood duration estimations. As
noticed in Section~\ref{subsec:likfun}, there is no finite
parametrization for the extremal dependence structure $V$ - see
Equation~\eqref{eq:jointDist}. Consequently, it seems reasonable to
suppose that one suited for flood hydrograph estimation may exist.

   \begin{figure}
     \centering
     \includegraphics[width=\textwidth]{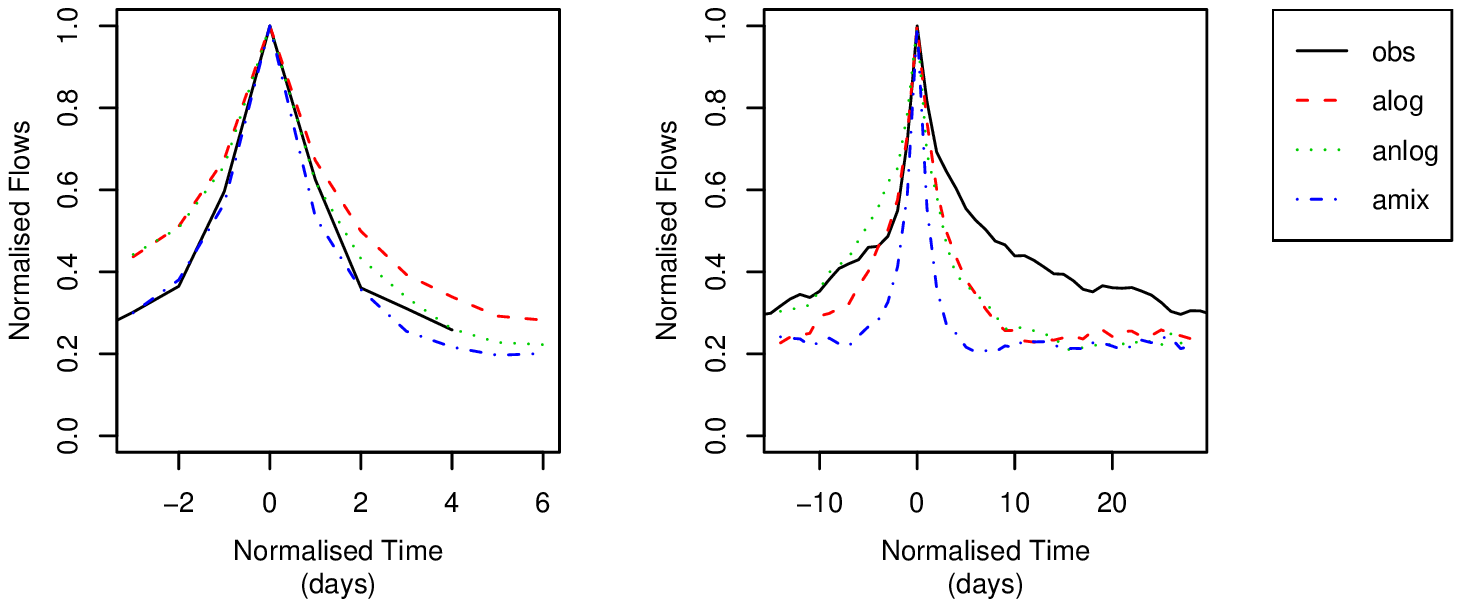}
     \caption{Observed and simulated normalized mean hydrographs for the
       J0621610 (left panel) and the L0400610 (right panel) stations.}
     \label{fig:longDurationPbm}
\end{figure}

Figure~\ref{fig:longDurationPbm} depicts the observed normalized mean
hydrographs and the ones predicted by the three asymmetric Markovian
models. For the J0621610 station (left panel), the normalized
hydrograph is well estimated by the three models; whereas for the
L0400610 station (right panel), the normalized hydrograph is poorly
predicted. This result confirms the inability of the three Markovian
models to reproduce long flood events with daily data and a first
order Markov chain.

   \begin{figure}
     \centering
     \includegraphics[angle=-90,width=\textwidth]{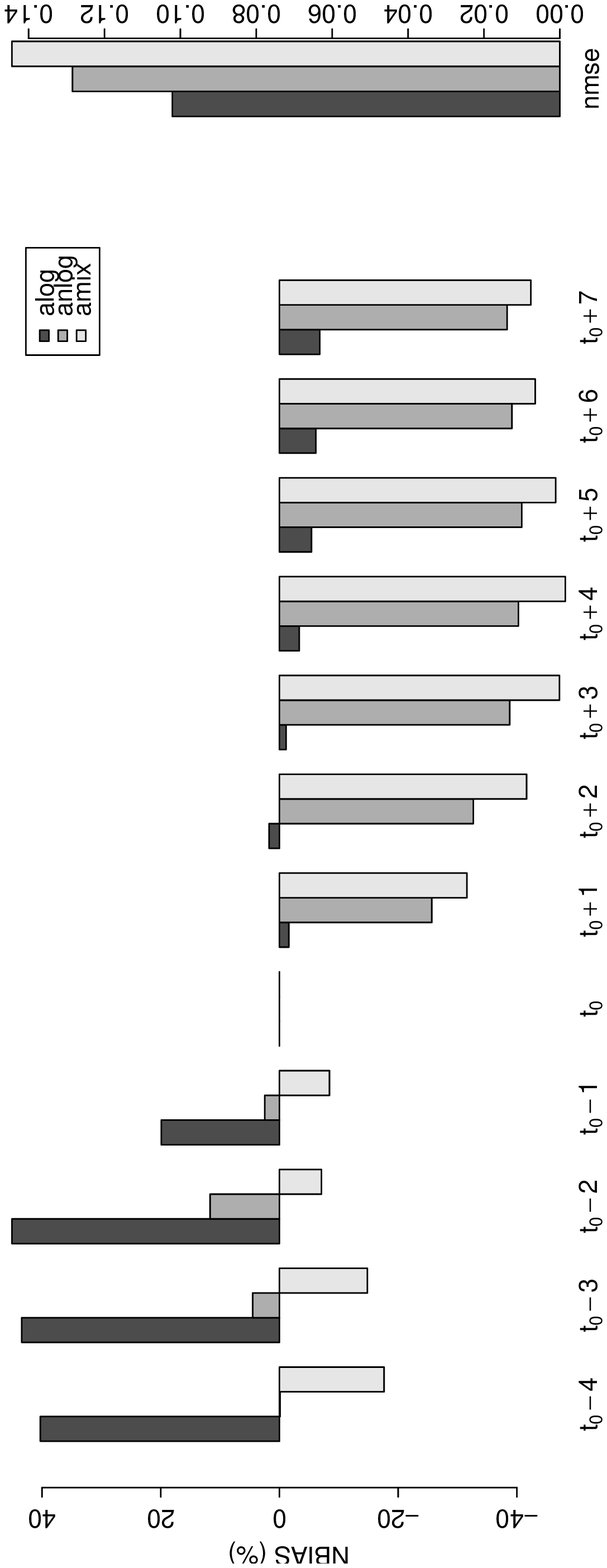}
     \caption{Evolution of the biases for the normalized mean hydrograph
       estimations in function of the distance of the flood peak time.}
     \label{fig:biasVsTime}
\end{figure}

Figure~\ref{fig:biasVsTime} represents the biases related to each
value of the normalized mean hydrograph. In addition, to help
estimator comparison, the $nmse$ is reported at the right side. It can
be seen that the $alog$ model dramatically overestimates the
hydrograph rising limb while giving reasonable estimations for the
falling phase. The $anlog$ model slightly overestimates the rising
part while strongly underestimates the falling one. The $amix$ model
always leads to underestimations - this is more pronounced for the
falling limb. However, despite these different behaviors, these three
estimators seems to have a similar performance - in terms of $nmse$.

   \begin{figure}
     \centering
     \includegraphics[width=\textwidth]{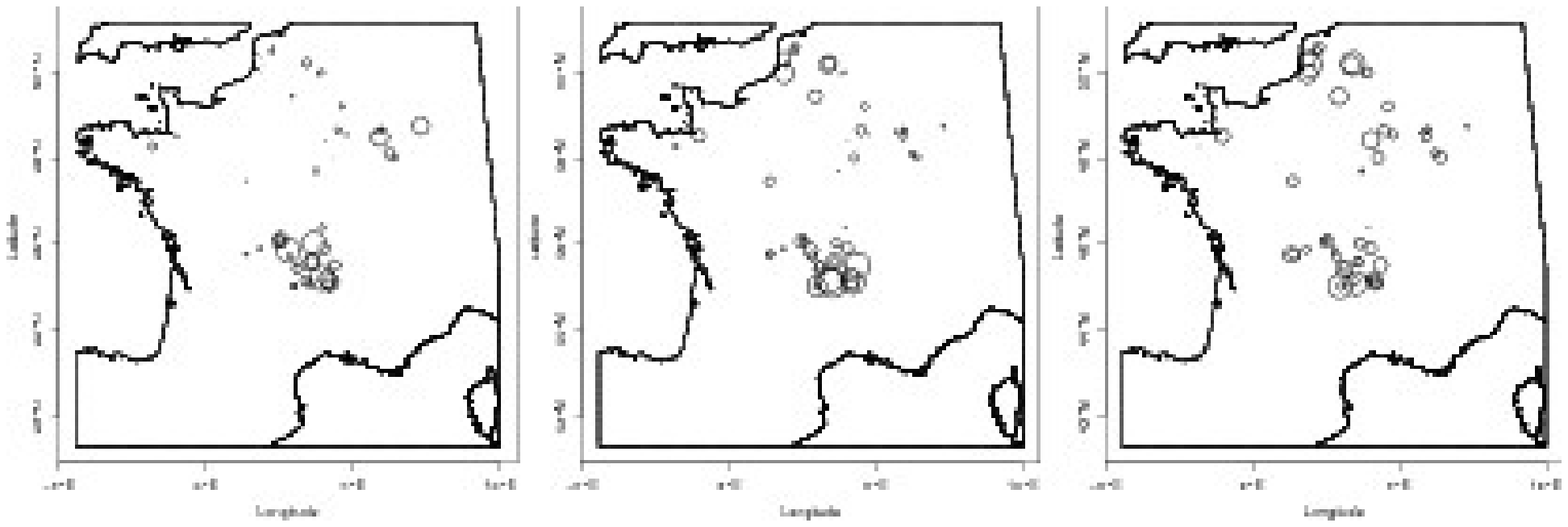}
     \caption{$nmse$ spatial distribution according for the three
       Markovian models. Left panel: $alog$, middle panel: $anlog$ and
       right panel: $amix$. The radius is proportional to the $nmse$
       value.}
     \label{fig:classFrance}
\end{figure}

Figure~\ref{fig:classFrance} represents the spatial distribution of
the $nmse$ on the normalized mean hydrograph estimation for each
Markovian model. It seems that there is a specific spatial
distribution. In particular, the worst cases are related to the middle
part of France. In addition, for different extremal dependence
structures, the best $nmse$ values correspond to different spatial
locations. The $alog$ model is more accurate for the extreme north
part of France; the $anlog$ model is more efficient for the east part
of France; while the $amix$ model performs best in the middle part of
France.  Consequently, as at a global scale no model is accurate to
estimate the normalized mean hydrograph, it is worth trying to
identify which catchment types are related to the best estimations.

For our data set, this is a considerable task. No standard statistical
technique lead to reasonable results. In particular, the principal
component analysis, hierarchical classification, sliced inverse
regression lead to no conclusion about which catchment types are more
suitable for our models. Only a regression approach gives some first
guidelines. For this purpose, a regression between the $nbias$ on the
$d_{mean}$ estimation for each asymmetric model and some
geomorphologic and hydrologic indices are performed. The effect of the
drainage area, an index of catchment slope derived from the
hypsometric curve \citep{Roche1963}, the Base Flow Index
($\mathbf{BFI}$) \citep[][Section 5.3.3]{Tallaksen2004} and an index
characterizing the rainfall persistence \citep{Vaskova1998} are
considered.

   \begin{eqnarray}
     \label{eq:nbias.mean.anlog}
     nbias\left(d_{mean};anlog\right) &=& 0.89 - 2.19 BFI, \qquad R^2 =
     0.40\\ 
     \label{eq:nbias.mean.amix}
     nbias\left(d_{mean};amix\right) &=& 0.49 - 1.74 BFI, \qquad R^2 =
     0.43
   \end{eqnarray}

   From equations~\eqref{eq:nbias.mean.anlog}
   and~\eqref{eq:nbias.mean.amix}, the $BFI$ variable explains around
   40\% of the variance. Despite the fact that a large variance
   proportion is not taken into account, the $BFI$ is clearly related
   to the $d_{mean}$ estimation performance. These equations indicate
   that the $anlog$ (resp. $amix$) model is more accurate to reproduce
   the $d_{mean}$ variable for gaging stations with a $BFI$ around 0.4
   (resp. 0.28).  These $BFI$ values correspond to catchments with
   moderate up to flash flow regimes respectively. These results
   corroborate the ones derived from Figure~\ref{fig:biasVsTime}: the
   first order Markovian models with a 1-day lag conditioning are not
   appropriate for long flood duration estimations. Consequently,
   while no physiographic characteristic is related to the $alog$
   performance; it is suggested, for such 1-day lag conditioning, to
   use the $anlog$ and $amix$ models for quick basins.

   \section{Conclusion}
   \label{sec:conclusion}

   Despite that univariate EVT is widely applied in environmental
   sciences, its multivariate extension is rarely considered. This
   work tries to promote the use of the MEVT in hydrology.  In this
   work, the bivariate case was considered as the dependence between
   two successive observations was modeled using a first order Markov
   chain. This approach has two main advantages for practitioners as:
   (a) the number of data to be inferred increases considerably and
   (b) other features can be estimated - flood duration, volume.

   In this study, a comparison between six different extremal
   dependence structures (including both symmetric and asymmetric
   forms) has been performed. Results show that an asymmetric
   dependence structure is more relevant. From a hydrological point of
   view, this asymmetry is rational as flood hydrographs are
   asymmetric. In particular, for our data, the asymmetric mixed model
   gives the most accurate flood peak estimations and clearly improves
   flood peak estimations compared to conventional estimators
   independently of the return period considered.

   The ability of these Markovian models to estimate the flood
   duration was carried out. It has been shown that, at first sight,
   no dependence structure is able to reproduce the flood hydrograph
   accurately. However, it seems that the $anlog$ and $amix$ models
   may be more appropriate when dealing with moderate up to flash flow
   regimes. These results depend strongly on the conditioning term
   (i.e.\ $\Pr[Y_t \leq y_t | Y_{t-\delta} = y_{t-\delta}]$) of the
   first order Markov chain and on the auto-correlation within the
   time series. In our application, $\delta = 1$ and daily time step
   was considered.


   More general conclusions can be drawn. The weakness of the proposed
   models to derive consistent flood hydrographs may not be related to
   the daily time step but to the inadequacy between the conditioning
   term and the flood dynamics. To ensure better results, higher order
   Markov chains may be of interest \citep{Fawcett2006}. However, as
   numerical problems may arise, another alternative may be to still
   consider a first order chain but to change the ``conditioning lag
   value'' $\delta$.  In particular, for some basins, it may be more
   relevant to condition the Markov chain with a larger but more
   appropriate lag value.


   Another option to improve the proposed models for flood hydrograph
   estimation is to use a more suitable dependence function $V$. As
   there is no finite parametrization for the extremal dependence
   structure, it seems reasonable that one more appropriate for flood
   hydrographs may exist. In this work, results show that the $anlog$
   model is more able to reproduce the hydrograph rising part, while
   the $alog$ is better for the falling phase. Define
   
   \begin{equation*}
     V\left(z_1,z_2\right) =  \alpha V_1\left(z_1,z_2\right) + \beta
     V_2\left(z_1,z_2\right)
\end{equation*}
   where $V_1$ (resp. $V_2$) is the extremal dependence function for the
   $alog$ (resp. $anlog$) model and $\alpha$ and $\beta$ are real
   constants such as $\alpha + \beta = 1$. By definition, $V$ is a new
   extremal dependence function. In particular, $V$ may combine the
   accuracy of the $alog$ and $anlog$ models for both the rising and
   falling part of the flood hydrograph. Another alternative may be to
   look at non-parametric Pickands' dependence function estimators
   \citep{Caperaa1997} but that will require techniques to simulate
   Markov chains from these non-parametric estimations.
   
   All statistical analysis were performed within the \citet{Rsoft}
   framework. In particular, the POT package \citep{Ribatet2007b}
   integrates the tools that were developed to carry out the modeling
   effort presented in this paper. This package is available, free of
   charge, at the website \url{http://www.R-project.org}, section
   CRAN, Packages or at its own webpage
   \url{http://pot.r-forge.r-project.org/}.

   \section*{Acknowledgments}
   \label{sec:acknowledgements}

   The authors wish to thank the French HYDRO database for providing
   the data. Benjamin Renard is acknowledged for criticizing
   thoroughly the data analyzed in this study.

   \appendix
   \section{Parametrization for the Extremal Dependence}
   \label{sec:paramExtDep}
   
   This annex presents some useful results for the six extremal
   dependence models that have been considered in this work. As first
   order Markov chains were used, only the bivariate results are
   described.
   
   \begin{table}
     \caption{Partial and mixed partial derivatives, definition domain,
       total independent and perfect dependent cases for each extremal
       symmetric dependence function $V$.}
     \label{tab:sumTabSym}
     \centering
     \begin{tabular}{lccc}
       \hline
       \multirow{2}*{Model} & \multicolumn{3}{c}{Symmetric Models}\\
       \cline{2-4}
       & $log$ & $nlog$ & $mix$\\
       \hline
       $V(x,y)$ & $\left(x^{-1/\alpha} + y^{-1/\alpha} \right)^\alpha$ &
       $\frac{1}{x} + \frac{1}{y} - \left(x^\alpha +
         y^\alpha\right)^{-1/\alpha}$ & $\frac{1}{x} + \frac{1}{y} -
       \frac{\alpha}{x+y}$\\
       $V_1(x,y)$ & $-x^{-\frac{1}{\alpha}-1}
       V(x,y)^{\frac{\alpha-1}{\alpha}}$ & $-\frac{1}{x^2} + x^{\alpha-1}
       \left(x^\alpha + y^\alpha\right)^{-\frac{1}{\alpha}-1}$ &
       $-\frac{1}{x^2} + \frac{\alpha}{\left(x+y\right)^2}$\\ 
       $V_2(x,y)$ & $-y^{-\frac{1}{\alpha}-1}
       V(x,y)^{\frac{\alpha-1}{\alpha}}$ & $-\frac{1}{y^2} + y^{\alpha-1}
       \left(x^\alpha + y^\alpha\right)^{-\frac{1}{\alpha}-1}$ &
       $-\frac{1}{y^2} + \frac{\alpha}{\left(x+y\right)^2}$\\
       $V_{12}(x,y)$ & $-(xy)^{-\frac{1}{\alpha}-1}
       \frac{1-\alpha}{\alpha} V(x,y)^{\frac{\alpha-2}{\alpha}}$ &
       $-(\alpha+1) (xy)^{\alpha-1} \left(x^\alpha +
         y^\alpha\right)^{-\frac{1}{\alpha}-2}$ &
       $-\frac{2\alpha}{\left(x+y\right)^3}$\\
       $A(w)$ & $\left[ (1-w)^{\frac{1}{\alpha}} + w^{\frac{1}{\alpha}}
       \right]^{\alpha}$ & $1 - \left[ (1 - w)^{-\alpha} + w^{-\alpha}
       \right]^{-\frac{1}{\alpha}}$ & $1 - w  \left(1 -w\right) \alpha$
       \\
       Independence & $\alpha = 1$ & $\alpha \rightarrow 0$ & $\alpha =
       0$\\
       Total dependence & $\alpha \rightarrow 0$ & $\alpha \rightarrow
       +\infty$ & Never reached\\
       Constraint & $0 < \alpha \leq 1$ & $\alpha > 0$ & $0\leq \alpha
       \leq 1$\\
       \hline
     \end{tabular}
\end{table}

   \begin{sidewaystable}
     \caption{Partial and mixed partial derivatives, definition domain,
       total independent and perfect dependent cases for each extremal
       asymmetric dependence function $V$.}
     \label{tab:sumTabAsym}
     \centering
     \begin{tabular}{lccc}
       \hline
       \multirow{2}*{Model} & \multicolumn{3}{c}{Asymmetric Models}\\
       \cline{2-4}
       & $alog$ & $anlog$ & $amix$\\
       \hline $V(x,y)$ & $\frac{1 - \theta_1}{x} + \frac{1 -
         \theta_2}{y} +
       \left[\left(\frac{x}{\theta_1}\right)^{-1/\alpha} +
         \left(\frac{y}{\theta_2}\right)^{-1/\alpha}\right]^\alpha$ &
       $\frac{1}{x} + \frac{1}{y} -
       \left[\left(\frac{x}{\theta_1}\right)^{\alpha} +
         \left(\frac{y}{\theta_2}\right)^\alpha\right]^{-1/\alpha}$ &
       $\frac{1}{x} + \frac{1}{y} - \frac{\left(2\alpha + \theta\right)
         x + \left(\alpha + \theta\right)y}{\left(x+y\right)^2}$\\
       $V_1(x,y)$ & $-\frac{1-\theta_1}{x^2} -
       \theta_1^{\frac{1}{\alpha}} x^{-\frac{1}{\alpha} - 1}
       \left[\left(\frac{x}{\theta_1}\right)^{-1/\alpha} +
         \left(\frac{y}{\theta_2}\right)^{-1/\alpha}\right]^{\alpha-1}$
       & $-\frac{1}{x^2} + \theta_1^{-\alpha} x^{\alpha-1}
       \left[\left(\frac{x}{\theta_1}\right)^{\alpha} +
         \left(\frac{y}{\theta_2}\right)^\alpha\right]^{-1/\alpha-1}$ &
       $-\frac{1}{x^2} - \frac{2\alpha + \theta}{(x+y)^2} + 2
       \frac{(2\alpha + \theta)x + (\alpha + \theta)y}{(x+y)^3}$\\
       $V_2(x,y)$ & $-\frac{1-\theta_2}{y^2} -
       \theta_2^{\frac{1}{\alpha}} y^{-\frac{1}{\alpha} - 1}
       \left[\left(\frac{x}{\theta_1}\right)^{-1/\alpha} +
         \left(\frac{y}{\theta_2}\right)^{-1/\alpha}\right]^{\alpha-1}$
       & $-\frac{1}{y^2} + \theta_2^{-\alpha} y^{\alpha-1}
       \left[\left(\frac{x}{\theta_1}\right)^{\alpha} +
         \left(\frac{y}{\theta_2}\right)^\alpha\right]^{-1/\alpha-1}$ &
       $-\frac{1}{y^2} - \frac{\alpha + \theta}{(x+y)^2} + 2
       \frac{(2\alpha + \theta)x + (\alpha + \theta)y}{(x+y)^3}$\\
       $V_{12}(x,y)$ & $\frac{\alpha - 1}{\alpha}
       (\theta_1\theta_2)^{\frac{1}{\alpha}} (xy)^{-\frac{1}{\alpha} -
         1} \left[\left(\frac{x}{\theta_1}\right)^{-1/\alpha} +
         \left(\frac{y}{\theta_2}\right)^{-1/\alpha}\right]^{\alpha-2}$
       & $-(\alpha+1) (\theta_1\theta_2)^{-\alpha} (xy)^{\alpha-1}
       \left[\left(\frac{x}{\theta_1}\right)^{\alpha} +
         \left(\frac{y}{\theta_2}\right)^\alpha\right]^{-1/\alpha-2}$ &
       $\frac{6\alpha + 4\theta}{(x+y)^3} - 6\frac{(2\alpha + \theta)x
         + (\alpha + \theta)y}{(x+y)^4}$\\
       $A(w)$ & $\left(1 - \theta_1 \right) \left(1 - w\right) +
       \left(1 - \theta_2 \right) w + \left[ (1 - w)^{\frac{1}{\alpha}}
         \theta_1^{\frac{1}{\alpha}} + w^{\frac{1}{\alpha}}
         \theta_2^{\frac{1}{\alpha}} \right]^{\alpha}$ & $1 - \left[
         \left(\frac{1 - w}{\theta_1} \right)^{-\alpha} +
         \left(\frac{w}{\theta_2} \right)^{-\alpha}
       \right]^{-\frac{1}{\alpha}}$ & $\theta w^3 + \alpha w^2 -
       \left(\alpha + \theta \right) w + 1$\\
       Independence & $\alpha = 1$ or $\theta_1 = 0$ or $\theta_2 = 0$
       & $\alpha \rightarrow 1$ or $\theta_1 \rightarrow 0$ or
       $\theta_2\rightarrow
       = 0$ & $\alpha = \theta = 0$\\
       Total dependence & $\alpha \rightarrow 0$ & $\alpha
       \rightarrow +\infty$ & Never reached\\
       Constraint & $0 < \alpha \leq 1$, $0 \leq \theta_1, \theta_2
       \leq 1$ & $\alpha >0$, $0 < \theta_1,\theta_2 \leq 1$ & $\alpha
       \geq 0$,
       $\alpha + 2 \theta \leq 1$, $\alpha + 3 \theta \geq 0$\\
       \hline
     \end{tabular}
\end{sidewaystable}
\end{doublespace}
\clearpage
\bibliography{/home/mathieu/Docs/LateX/biblio_ribatet}
\bibliographystyle{plainnat}
\end{document}